\documentclass[twocolumn]{aastex63}
\usepackage{lineno}
\usepackage{amsmath}
\usepackage{amssymb}
\usepackage{epsfig}
\usepackage{natbib}
\usepackage{graphicx}
\usepackage{color}
\usepackage{multirow}
\usepackage{subfigure}
\usepackage{ulem}
\usepackage{url}
\def\UrlBreaks{\do\A\do\B\do\C\do\D\do\E\do\F\do\G\do\H\do\I\do\J
\do\K\do\L\do\M\do\N\do\O\do\P\do\Q\do\R\do\S\do\T\do\U\do\V
\do\W\do\X\do\Y\do\Z\do\[\do\\\do\]\do\^\do\_\do\`\do\a\do\b
\do\c\do\d\do\e\do\f\do\g\do\h\do\i\do\j\do\k\do\l\do\m\do\n
\do\o\do\p\do\q\do\r\do\s\do\t\do\u\do\v\do\w\do\x\do\y\do\z
\do\.\do\@\do\\\do\/\do\!\do\_\do\|\do\;\do\>\do\]\do\)\do\,
\do\?\do\'\do+\do\=\do\#}

\newcommand       \G            {{\rm G}}
\newcommand       \GBP       {G_{\rm BP}}
\newcommand       \GRP       {G_{\rm RP}}
\newcommand       \mum        {\,{\rm \mu m}}
\newcommand       \Ks           {{ K_{\rm S}}}
\newcommand       \K             {\,{\rm K}}
\newcommand       \Teff           {{ T_{\rm eff}}}

\newcommand{\AV}{A_{\rm V}}
\newcommand{\Av}{A_{\rm V}}
\newcommand{\AB}{A_{\rm B}}

\newcommand{\ABP}{A_{G_{\rm BP}}}
\newcommand{\ARP}{A_{G_{\rm RP}}}
\newcommand{\RV}{R_{\rm V}}
\newcommand{\Rv}{R_{\rm V}}

\newcommand{\MRP}{M_{G_{\rm RP}}}

\shorttitle{LMC \& SMC Extinction Law}
\shortauthors{Wang \& Chen}

\begin{document}

\title{The Optical to Infrared Extinction Law of Magellanic Clouds Based on Red Supergiant and Classical Cepheid}

\correspondingauthor{Shu Wang, Xiaodian Chen}
\email{shuwang@nao.cas.cn, chenxiaodian@nao.cas.cn}

\author[0000-0003-4489-9794]{Shu Wang}
\affiliation{CAS Key Laboratory of Optical Astronomy, National Astronomical Observatories, 
Chinese Academy of Sciences, Beijing 100101, China}
\affiliation{Department of Astronomy, China West Normal University, Nanchong, China}

\author[0000-0001-7084-0484]{Xiaodian Chen}
\affiliation{CAS Key Laboratory of Optical Astronomy, National Astronomical Observatories, 
Chinese Academy of Sciences, Beijing 100101, China}
\affiliation{Department of Astronomy, China West Normal University, Nanchong, China} 
\affiliation{School of Astronomy and Space Science, University of the Chinese Academy of Sciences, Beijing 101408, China} 
\affiliation{Institute for Frontiers in Astronomy and Astrophysics, Beijing Normal University,  Beijing 102206, China}

\begin{abstract}
Precise interstellar dust extinction laws are important to infer the intrinsic properties of reddened objects and correctly interpret observations. 
In this work, we attempt to measure the optical--infrared extinction laws of the Large and Small Magellanic Clouds (LMC and SMC) by using red supergiant (RSG) stars and classical Cepheids as extinction tracers. 
The spectroscopic RSG samples are constructed based on the APOGEE spectral parameters, Gaia astrometric data, and multi-band photometry.
We establish the effective temperature--intrinsic color relations for RSG stars and determine the color excess ratio (CER) $E(\lambda - \GRP)/E(\GBP - \GRP)$ for LMC and SMC. 
We use classical Cepheids to derive base relative extinction $\ARP/E(\GBP-\GRP)$. The results are $1.589\pm0.014$ and $1.412\pm0.041$ for LMC and SMC. 
By combining CERs with $\ARP/E(\GBP-\GRP)$, the optical--infrared extinction coefficients $A_\lambda/\ARP$ are determined for 16 bands. 
We adjust the parameters of $\RV$-dependent extinction laws and obtain the average extinction laws of LMC and SMC as $\RV=3.40\pm0.07$ and $\Rv=2.53\pm0.10$, which are consistent with \citet{2003ApJ...594..279G}. In the optical bands, the adjusted $\Rv$ extinction curves agree with the observations with deviations less than 3\%.  
\end{abstract}


\section{Introduction}\label{intro}

The interstellar extinction curve is the absorption and scattering of starlight by dust as a function of wavelength. 
It is crucial to correct the effects of dust extinction and infer the intrinsic properties of reddened objects. 
Besides, the extinction curve contains important information about the interstellar dust grains, such as the size distribution and composition. Accurate extinction curves are also essential in distance determination and galaxy structure studies. 
Current studies of interstellar extinction are mainly concentrated on the Milky Way. 
The Magellanic Clouds (MCs), including the Large Magellanic Cloud (LMC) and Small Magellanic Cloud (SMC), are satellite galaxies of the Milky Way. 
They are low-metallicity dwarf galaxies and offer a unique opportunity to study of the dust extinction properties in diverse galactic environments.

However, compared to the Milky Way, the studies of the LMC and SMC extinction curves are limited. 
In the last two decades, \citet{2003ApJ...594..279G} measured ultraviolet (UV) to near-infrared (IR) extinction curves of 24 sightlines in the MCs (19 in the LMC and five in the SMC). 
\citet{2013ApJ...776....7G} investigated the mid-IR extinction law and its variation in the LMC.  
\citet{2014AA...564A..63M} and \citet{2016MNRAS.455.4373D} studied the extinction law of the Tarantula Nebula (30 Doradus) in the LMC at different bands based on different data and methods, respectively. 
\citet{2012A&A...541A..54M} calculated the UV extinction law for four stars in a quiescent cloud in the SMC.
\citet{2017ApJ...847..102Y} measured the extinction curve of a $\sim 200\ {\rm pc} \times 100\ {\rm pc}$ region in the southwest bar of the SMC.  
\citet{2017MNRAS.466.4540H} analyzed the UV extinction curve and its regional variation in the SMC. 
These studies have focused on limited sightlines of O- and B-type stars or on particular regions. 
Hence, it is necessary to study the extinction law covering a wider area of the LMC and SMC. 

To measure the extinction curves, two methods are commonly used. 
One is the ``pair method'', which is widely applied to spectroscopic data or the spectral energy distribution (SED) of multi-band photometric data. The other is the color-excess method for photometric data. 
The ``pair method'' calculates the extinction curve by comparing the observed spectrum (or SED) of a reddened star with that of an unreddened star (or the theoretical stellar spectrum) of the same spectral type. 
O- and B-type stars are often taken as extinction tracers as they are very bright and easy to observe \citep[e.g.,][]{1990ApJS...72..163F, 2007ApJ...663..320F, 2003ApJ...594..279G, 2012A&A...541A..54M, 2014AA...564A..63M, 2019ApJ...886..108F,2021ApJ...916...33G}. 
However, the small number of these stars and their location in atypical environments limit their widespread application. 
Besides, the selection of spectral templates and the zero point at spectral comparison can lead to systematic errors in extinction measurements. 
The color-excess method computes the ratio of two color excesses (CEs), $E(\lambda-\lambda_1)/E(\lambda_1-\lambda_2)$, for a group of stars based on the linear fitting of CE--CE diagrams. With the known distance, the total-to-selective extinction ratio, $A_\lambda/E(\lambda-\lambda_1)$, can be derived from the slope of the reddening vector on the color-magnitude diagrams \citep[CMDs,][]{2009ApJ...696.1407N, 2016MNRAS.455.4373D, 2017ApJ...847..102Y}. 
Red giant branch (RGB) stars and red clump (RC) stars are appropriate tracers because their intrinsic color indices are relatively consistent or easy to measure \citep[e.g.,][]{2005ApJ...619..931I, 2007ApJ...663.1069F, 2009ApJ...707..510Z, 2009ApJ...707...89G, 2013ApJ...776....7G, 2013ApJ...773...30W, 2017ApJ...848..106W, 2019ApJ...877..116W}. 
RGB stars and RC stars are numerous and very bright in the IR, and they are good tracers of the IR interstellar extinction. 
However, the contamination by other types of stars is a problem. 
Moreover, the adoption of constant intrinsic color indices of RGB stars and RC stars also introduces errors in extinction estimates, because they are closely related to stellar parameters such as effective temperature, metallicity, and surface gravity. 
Furthermore, the absolute magnitude of RC stars is also not constant, but related to the metallicity and age \citep{2021ApJ...923..145W}. 
These problems can be effectively solved only when the spectral parameters of the tracers are available.

In summary, stars used as extinction tracers have the following characteristics: 1) they are very luminous, which allows us to detect distant or high extinction regions; 2) The intrinsic color index is almost invariant, or can be inferred from their stellar parameters.
In this work, we propose to use red supergiant (RSG) stars and classical Cepheids as tracers to determine the dust extinction. 
RSG stars are massive stars with masses ranging from $\sim$\ 8 to 30\ $M_\odot$. 
Typically, RSG stars have high luminosities of $\sim 4000-40 0000\ L_\odot$, 
low effective temperatures of $\sim3300-5000\K$, and radii up to $1500\ R_\odot$ \citep{2005ApJ...628..973L, 2007ApJ...660..301M, 2012A&A...540L..12W, 2013ApJ...767....3D, 2018A&A...616A.175Y, 2021A&A...646A.141Y, 2021ApJ...923..232R}.  
They are the descendants of H-burning O- and B-type massive stars and thought to evolve horizontally on the Hertzsprung-Russell (H-R) diagram. 
The main advantage of using RSG stars as extinction tracers is that they are $4-8$ magnitude brighter than RGB stars or RC stars. Besides, RSG stars are very young and often present near dusty clouds. 
Thus, RSG stars can be used to study the extinction of high extinction regions or even of other galaxies in the Local Group. 
Moreover, the intrinsic color indices of RSG stars can be well estimated from their stellar parameters.

In this work, we explore whether RSG stars can be used as tracers of interstellar extinction and combine them with classical Cepheids to study dust extinction in the LMC and SMC.  
Based on the APOGEE \citep[the Apache Point Observatory Galactic Evolution Experiment,][]{2011AJ....142...72E,2017AJ....154...94M} survey data, we select a group of RSG stars with high-quality stellar atmospheric parameters and chemical abundances. 
We establish the effective temperature $\Teff$--intrinsic color relations of RSG stars to obtain CE values.  
The color excess ratios (CERs) are determined by using the color-excess method based on RSG stars. 
With the help of classical Cepheids, we derive the base relative extinction $\ARP/E(\GBP - \GRP)$.  
After that , we convert the CERs into the relative extinction $A_{\lambda}/\ARP$ and obtain the optical to IR dust extinction curves of the LMC and SMC. 
In addition, we discuss the effects of circumstellar dust and variability of RSG stars on their adoption as interstellar extinction tracers.

The sketch of this paper is as follows. 
The description of data sets and the construction of the RSG sample are presented in Section~\ref{data}. 
We also provide a catalog of spectroscopic RSG stars in Section~\ref{data}. 
In Section~\ref{colorexcess}, we describe the method to derive the relation between $\Teff$ and intrinsic color indices for RSG stars. The determination of CERs and relative extinction is also in Section~\ref{colorexcess}. In Section~\ref{extlaw}, we analyze the optical to IR extinction and reddening curves of the LMC and SMC by comparing the observations with the models. 
Then we adjust the $\RV$-dependent extinction laws. 
We compare our extinction law with previous works in Section~\ref{discussion}. 
We also discuss the effects of the possible presence of circumstellar dust around RSG stars and the variability of RSG stars on the determination of LMC and SMC interstellar extinction in Section~\ref{discussion}.  
We summarize our main conclusions in Section~\ref{conclusion}.

\section{Data and Sample} \label{data}
\subsection{Data} \label{data1}

To determine the extinction law of MCs, we collected spectroscopic data from APOGEE, and optical to near-IR photometric data from the Gaia, MCPS \citep[Magellanic Clouds Photometric Survey,][]{2002AJ....123..855Z, 2004AJ....128.1606Z}, SMSS \citep[SkyMapper Southern Survey,][]{2018PASA...35...10W}, SMASH \citep[Survey of the Magellanic Stellar History,][]{2017AJ....154..199N}, and 2MASS \citep[Two Micron All Sky Survey,][]{2006AJ....131.1163S} surveys.

\subsubsection{Gaia}

The third Gaia data release, Gaia DR3, contains a full astrometric solution for about 1.46 billion sources brighter than 21.0 mag in $G$ band \citep{2022arXiv220800211G}. 
It also provides broad-band photometry in the $G$, $\GBP$, $\GRP$ bands for more than 1.5 billion sources and the radial velocity measurements for about 33 million bright stars \citep{2022arXiv220800211G}. 
The astrometric data in Gaia DR3 are the same as those of Gaia early DR3. 
The typical uncertainties for proper motion and mean $G$-band photometry are $0.02-0.04$ mas yr$^{-1}$ at $G < 15$, 1 mmag at $G=17$ \citep{2021A&A...649A...1G}.
In this work, we use the Gaia proper motion and radial velocity to exclude foreground stars. 
Then, we use Gaia photometric data to determine the extinction law of MCs.

\subsubsection{APOGEE} 
APOGEE is a high-resolution ($R\sim22,500$) near-IR $H$-band ($15000-17000{\rm \AA}$) spectroscopic survey \citep{2017AJ....154...94M}. 
It observed the sky by using the 2.5-meter Sloan Telescope at the Apache Point Observatory and the 2.5-meter du Pont Telescope at the Las Campanas Observatory \citep{2006AJ....131.2332G, 2019PASP..131e5001W}. 
We used data from the latest released DR17 \citep{2022ApJS..259...35A}. 
DR17 is a component of the fourth phase of Sloan Digital Sky Surveys \citep[SDSS-IV,][]{2017AJ....154...28B} and contains data and information for 657,135 unique targets in the Milky Way and nearby satellite galaxies. 
It provides the derived stellar properties, including radial velocities, stellar atmospheric parameters (e.g, $\Teff$, surface gravity $\log g$, and metallicity [M/H]), and individual chemical abundances (e.g. C, N, O, other $\alpha$, iron-peak elements). 
Typical internal uncertainties in the stellar atmospheric parameters are 27 K in $\Teff$, 0.04 dex in $\log g$, and 0.01 dex in [M/H].

\subsubsection{MCPS} 
The MCPS survey observed the central 18 deg$^2$ area of the SMC \citep{2002AJ....123..855Z} and 64 deg$^2$ area of the LMC \citep{2004AJ....128.1606Z} by using the Las Campanas Swope 1-meter telescope. 
It provides an MCPS catalog with $UBVI$ photometric data. 
In the MCPS catalog, for the brightest 0.02\% stars with $B$ or $V <13.5$ mag, their $U, B, V$ photometry has been replaced by Massey's catalog \citep[$UBVR$ bright star survey of the LMC and SMC,][]{2002ApJS..141...81M}. 
Note that Massey's catalog only covers 14.5 deg$^2$ of the LMC, a quarter of the area covered by the MCPS catalog, so most stars in the MCPS catalog with $B$ or $V <13.5$ mag still have suspect photometry after replacing the photometry with Massey's catalog.  
For SMC stars, the $I$-band photometry has been replaced by the OGLE photometry \citep[Optical Gravitational Lensing Experiment,][]{1998AcA....48..147U} for bright stars with $I < 13.5$ mag. The photometric uncertainties of stars with good quality should be less than 0.4 mag in the $U$ band and 0.2 mag in the $B, V, I$ bands \citep{2002AJ....123..855Z}.

\subsubsection{SMSS} 
SMSS is a southern hemisphere image survey using the dedicated 1.3-meter SkyMapper telescope at Siding Spring Observatory in Australia \citep{2018PASA...35...10W}. 
The DR1 catalog presents data from the shallow survey including objects with $\sim8-18$ mag in all six bands: $u$, $v$, $g$, $r$, $i$, and $z$ \citep{2018PASA...35...10W}. 
The latest DR2 includes a part of the main survey (100 second exposures), with the limit magnitudes reaching 21st magnitude in the $g$ and $r$ bands \citep{2019PASA...36...33O}. 
By the end of the survey, the photometry depths can reach $20-22$ mag in all six bands \citep{2019PASA...36...33O}. 
We use the data published in DR2 which has a precision of 1.0\% in $u$ and $v$ bands, and 0.7\% in $g, r, i, z$ bands as measured by internal reproducibility \citep{2019PASA...36...33O}.

\subsubsection{SMASH} 
The SMASH survey is an NOAO survey project using the NOAO Blanco 4-meter telescope located at Cerro Tololo Inter-American Observatory \citep{2017AJ....154..199N}. 
It observed about 480 deg$^2$ of the Magellanic system (the MC main bodies and the Magellanic periphery) with deep $u, g, r, i, z$ images \citep{2017AJ....154..199N}. 
Depending on the exposure times, the photometric coverage ranges from $\sim16$ mag (shallow, $\sim$ 60 s) to $24-25$ mag (deep, $\sim$ 300 s). 
We use the latest DR2 data with the photometric precision of 0.10, 0.07, 0.05, 0.08, and 0.05 mag in $u, g, r, i$, and $z$ band, respectively \citep{2021AJ....161...74N}. 
For the brightest stars, the completeness decreases. 
Hence, this work only uses $u$ and $g$ bands.

\subsubsection{2MASS} 
2MASS is an all-sky survey in the near-IR $J$, $H$, and $\Ks$ bands by using two 1.3-meter diameter telescopes located at Mount Hopkins, Arizona, and Cerro Tololo, Chile \citep{2006AJ....131.1163S}. 
The $10\sigma$ detection depths of the point source are 15.8, 15.1, and 14.3 mag in the $J$, $H$, and $\Ks$ bands, respectively \citep{2006AJ....131.1163S}. 
We use data from the 2MASS all-sky catalog of point sources \citep{2003yCat.2246....0C}.

\begin{figure*}[ht]
\centering
\vspace{-0.0in}
\includegraphics[angle=0,width=7.0in]{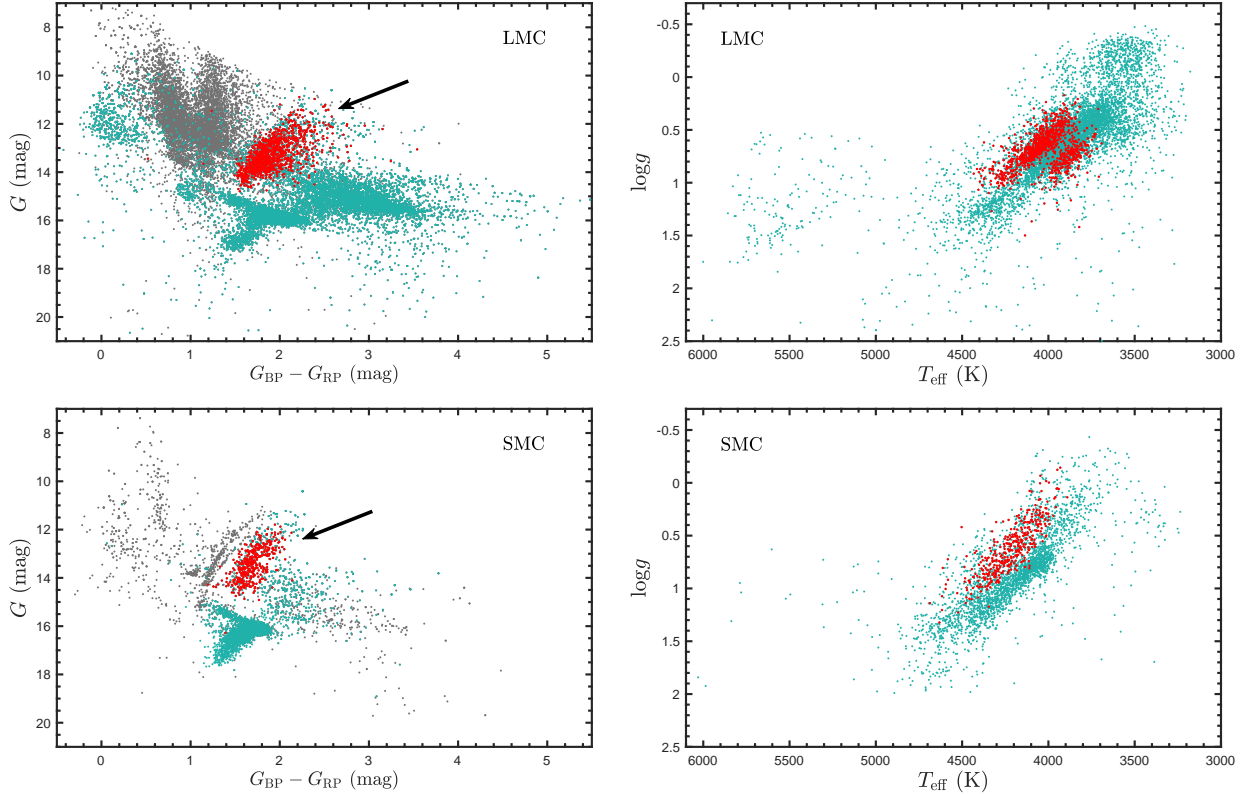}
\vspace{-0.2in}
\caption{\footnotesize
               \label{fig:sample}
The distribution of RSG stars (red dots) in the Gaia $(\GBP-\GRP)$ vs. $G$ CMD (left) and the $\Teff$ vs. $\log g$ diagram (right). The gray dots are the foreground Galactic sources, the turquoise dots are the LMC (top) and SMC (bottom) sources.} The arrows indicate the approximate positions of RSG candidates on CMDs.
\end{figure*}

\begin{table*}[ht]
\begin{center}
\caption{\label{tab:sample} Spectroscopic RSG Stars in the Magellanic Clouds}
\vspace{0.1in} 
\hspace{-1.0in}    
\begin{tabular}{lcccccccccccc}
\hline \hline     
ID & R.A.(J2000) & Decl.(J2000) & $\Teff$ &  $\log g$  &  [M/H] & $E(\GBP - \GRP)$ & $G$ & $\GBP$ &...$^*$ & $H_{\rm err}$ & $\Ks_{\rm err}$ & Galaxy\\   
 & (deg)  & (deg)  & (K) & & (dex) & (mag) & (mag) & (mag) &... & (mag) & (mag) &  \\
\hline                                                                                                   
1 & 76.863813 & -67.209457 & 4015.59 & 0.61 & -0.449 & 0.0528 & 13.792 & 14.722 & ... & 0.024 & 0.021 & LMC \\ 
2 & 83.260461 & -71.646545 & 4317.06 & 0.80 & -0.470 & 0.0544 & 13.877 & 14.637 & ... & 0.028 & 0.026 & LMC \\ 
3 & 74.098496 & -69.703072 & 3913.49 & 0.75 & -0.546 & 0.1831 & 12.048 & 13.160 & ... & 0.038 & 0.023 & LMC \\ 
4 & 82.024431 & -67.630287 & 3834.46 & 0.65 & -0.548 & 0.0779 & 12.010 & 13.140 & ... & 0.027 & 0.021 & LMC \\ 
5 & 82.823584 & -70.852493 & 4108.01 & 0.53 & -0.586 & 0.0486 & 12.913 & 13.771 & ... & 0.027 & 0.023 & LMC \\
... & ... & ... & ... &... & ... &... & ... &... &... & ... & ... & ... \\
\hline 
\end{tabular}
\end{center}
\tablenotetext{}{Note $*$: This table is available in its entirety in machine-readable form. 
The stellar parameters of $\Teff, \log g$ and [M/H] are taken from the APOGEE catalog.}         
\end{table*}

\subsection{The Spectroscopic RSG Catalog}\label{sample}

We construct a spectroscopic RSG sample mainly based on astrometric data from Gaia and spectroscopic data from APOGEE.
First, we use the Gaia parameters to exclude the foreground Galactic stars. 
The specific constraints are as follows: 
\begin{enumerate}
\item the location, $64^\circ<{\rm R.A.}<94^\circ$, $-74^\circ<{\rm Decl.}<-63^\circ$ for LMC and $3^\circ<{\rm R.A.}<25^\circ$, $-76^\circ<{\rm Decl.}<-70^\circ$ for SMC;  
\item the proper motion (PM) for LMC and SMC, $0.0\ {\rm mas\ yr}^{-1}<{\rm PM_{R.A.}}<2.5\ {\rm mas\ yr}^{-1}$ and $-2\ {\rm mas\ yr}^{-1}<{\rm PM_{Decl.}}<2\ {\rm mas\ yr}^{-1}$;  
\item the radial velocity (RV), $\rm RV>200\ km\ s^{-1}$ (LMC) and $\rm RV>100\ km\ s^{-1}$ (SMC);  
\item remove stars with parallaxes satisfy both $\varpi>0$ and $\,| \varpi-0.02\,| > 3\times\varpi_{\rm err}$ for LMC, satisfy both $\varpi>0$ and $\,| \varpi-0.0166 \,| > 3\times\varpi_{\rm err}$ for SMC.  
\end{enumerate}

The left panel of Figure~\ref{fig:sample} displays the selection of the RSG sample in the Gaia $(\GBP-\GRP)$ vs. $G$ CMD. 
The gray dots are the foreground Galactic sources, while the turquoise dots are the LMC (top) and SMC (bottom) sources.
The LMC and SMC sources mainly include RSG stars, asymptotic giant branch (AGB) stars, and RGB stars.
We take the LMC CMD diagram (top left panel of Figure~\ref{fig:sample}) as an example. 
The interstellar extinction causes the position of the star to shift from the upper left to the lower right, as does the broadening of RSG stars (red dots). Overall, the LMC extinction is not significant. 
The turquoise dots distributed at the lower right of the red dots are mainly AGB stars, which are elongated in the color $\GBP-\GRP$. This elongation is mainly caused by the different temperatures and the different circumstellar dust. 
The turquoise dots distributed below and parallel to the red dots are mainly RGB stars, which can also be used to measure the extinction law, but with a larger uncertainty, as discussed in Section~\ref{RGBext}.

The right panel of Figure~\ref{fig:sample} illustrates the distribution of LMC (top) and SMC (bottom) stars (turquoise dots) in the $\Teff$ vs. $\log g$ diagram. 
We further selected RSG stars using the APOGEE parameters. 
For this purpose, we analyzed the distribution of the APOGEE parameters of the RSG candidates, where the arrows in Figure~\ref{fig:sample} indicate their approximate positions on the CMDs \citep[see also Fig. 9 of ][]{2021A&A...646A.141Y}.   
Finally, the following APOGEE parametric and photometric criteria were determined to select RSG stars: 
\begin{enumerate}
\item the surface gravity $\log g < 2$; 
\item metallicity $-0.6\,{\rm dex}<[{\rm M/H}]<-0.4\,{\rm dex}$ (LMC) and $-1.0\,{\rm dex}<[{\rm M/H}]<-0.7\,{\rm dex}$ (SMC); 
\item $\alpha$-elemental abundance $-0.06\,{\rm dex}<[\alpha/{\rm M}]<0.02\,{\rm dex}$ (LMC) and $-0.12\,{\rm dex}<[\alpha/{\rm M}]<0.00\,{\rm dex}$ (SMC);
\item $\Ks<12.0$ mag. 
\end{enumerate}

After applying the above constraints, substantial RGB and AGB stars were removed. 
We further excluded possible AGB stars based on multi-band CMDs.   
Since AGB stars have larger observed color indices and lower $\Teff$ than RSG stars, we remove them if they lie in the red sequences of both the Gaia $(\GBP-\GRP)$ vs. $G$ and the 2MASS $(J-\Ks)$ vs. $\Ks$ CMDs. 
Meanwhile, we check whether these ejected stars are located at the low-temperature end of the $\Teff$ vs. $\log g$ diagram.
This check prevents us from removing RSG stars with high extinction.
Our final RSG sample contains 1,073 stars (LMC) and 398 stars (SMC), shown as red dots in Figure~\ref{fig:sample}.
For the LMC RSG sample, most stars are in $3700\K \le \Teff \le 4500\K$ and $0.3 < \log g < 1.0$.
For the SMC RSG sample, most stars are in $3900\K \le \Teff \le 4650\K$ and $0.2 < \log g < 1.0$.

We adopted a radius of 1$''$ to cross-matched the spectroscopic RSG sample with photometric catalogs listed in Section~\ref{data1}. 
The RSG stars are in the one-to-one match. When crossed with Gaia, there are two RSG stars with double-matched objects. In the $G$-band, the brightness ratios of faint and bright stars are 1\% and 8\%. If the angular resolution of other bands is worse than 1$''$, the faint stars cannot be distinguished, resulting in a blending effect. If the temperature of the faint star is significantly different from that of the RSG star, this will lead to a bias in our determined intrinsic color and CE. Here, only two stars have a weak blending problem, and the effect on the results is completely negligible. 
For the 2MASS bands, the angular resolution is around 2$''$. Using a 2$''$ radius crossed with Gaia, we found 169 stars with blending problems. However, the vast majority have a small effect on the IR luminosity, with an overall effect of 0.3\%. The effect on the IR CE is less than 0.002 mag, which is also negligible. \citet{2018A&A...616A.175Y} also confirmed that there is no significant blending for near-IR and mid-IR photometry in selecting RSGs candidates.

We present a catalog of spectroscopic RSG stars, including 1073 stars in LMC and 398 stars in SMC. 
This catalog gives the positions of the RSG stars (R.A., Decl.), the stellar parameters ($\Teff, \log g$, and [M/H]), the multi-band photometry with uncertainties (Gaia, MCPS, SMASH, SMSS, and 2MASS), and the derived CE $E(\GBP - \GRP)$ \footnote{For RSG stars in our catalog, 89 stars in the LMC and 26 stars in the SMC do not have $E(\GBP - \GRP)$ values. These stars were not used to study the extinction because they were excluded by either $\Ks-W2>0.13$ mag or the photometric uncertainties in $\GBP$ and $\GRP$ bands $>0.03$ mag.}. 
A portion of the catalog is shown in Table~\ref{tab:sample} for guidance regarding its form and content.

\citet{2003AJ....126.2867M} provided a catalog of 118 red stars toward the SMC and 167 red stars toward the LMC, of which 89\% and 95\% are RSG stars, respectively. 
After cross-matching with our catalog by a radius of 1$''$, we found that 45 RSG stars (28\%) of LMC and 36 RSG stars (34\%) of SMC are in our sample. 
The main difference between our RSG sample and Massey's sample is that we chose a sample with a more concentrated metallicity. We found that APOGEE data are available for 112 (LMC) and 80 (SMC) RSG stars in Massey's sample.
After applying our [M/H] and [$\alpha$/M] criteria, 49 (LMC) and 44 (SMC) remain. 
Then, after further removing possible contaminants according to multi-band CMDs and $\Teff$ vs. $\log g$ diagram, 45 (LMC) and 36 (SMC) are finally left. 
The metallicities of our RSG sample ($\sim1000$ RSG stars) are concentrated in the 0.2 dex range, while the metallicity of Massey's sample has a large dispersion, with about half of the RSG stars having metallicities below the lower limit of our selection criterion. For these low-metallicity RSG stars, the calculation of the intrinsic color would be less accurate. In addition, Massey's sample consists of the brightest RSG stars, and most of them have more circumstellar dust, so we did not include Massey's sample in this work.

\citet{2019A&A...629A..91Y, 2021A&A...646A.141Y} and \citet{2021ApJ...923..232R} also provided the LMC and SMC RSG catalogs.  
After integrating their catalogs and cross-matching them with APOGEE data, we obtained a combined catalog of 1846 (LMC) and 626 (SMC) RSG stars. 
We found 1040 (97\%) and 392 (99\%) of our RSG stars are in the combined catalog. 
This indicates that our selected RSG stars are reliable.

\subsection{The RSG Sample for Measuring Extinction}\label{extsample}

To study interstellar extinction with our selected RSG stars, we also need to remove sources with circumstellar dust. 
Evolved RSG stars may have a dusty circumstellar envelope due to mass loss. 
Dust absorbs UV/optical photons and radiates in the IR bands. 
If there is dust around RSG stars, it may lead to IR CE. 
Therefore, IR bands are often used to indicate stars with circumstellar envelopes. 
For example, \citet{2007ApJ...663.1069F} adopted mid-IR colors to exclude sources with circumstellar dust, such as pre-main-sequence stars, AGB stars, and young stellar objects. 
IR colors can be used to detect circumstellar dust tied to RSG stars 
\citep{2005A&A...438..273V, 2010AJ....140..416B, 2011AJ....142..103B, 2018A&A...616A.175Y}.  
We also tried to use IR colors to remove RSG stars with circumstellar dust. 
After checking multi-band colors, we used $\Ks$ minus WISE $W2$ (centered at $4.5\mum$) larger than 0.13 mag to remove stars with circumstellar envelope. 
71 and 14 stars belonging to the LMC and SMC were excluded.
More discussion on the effects of circumstellar dust can be found in Section~\ref{dust}.

Finally, to guarantee the photometric precision, we require the photometric uncertainties $\sigma_m$ of each catalog satisfy the followings:  
\begin{enumerate}
\item For MCPS data, $\sigma_m \le 0.4$ mag in $U$ band, and $\sigma_m \le 0.2$ mag in $B$, $V$, $I$ bands. 
\item For Gaia data, $\sigma_m \le 0.03$ mag in $G$, $\GBP$, $\GRP$ bands. 
\item For SMASH data, $\sigma_m \le 0.05$ mag in $u, g$ bands. 
\item For SMSS data, $\rm x\_flags<4$, $\rm x\_nimaflags<5$, $\rm x\_ngood > 0$ in ${\rm x}: v, g, r, i, z$ bands. 
\item For 2MASS data, $\sigma_m \le 0.05$ mag in $J$, $H$, $\Ks$ bands. 
\end{enumerate}

\section{The Optical to IR Extinction Curves} \label{colorexcess}

In this section, we describe the method and procedure for determining the wavelength-dependent extinction law. 
First, we establish the $\Teff$--$(\lambda - \GRP)_0$ relations to estimate the $(\lambda - \GRP)_0$ of each RSG star. 
Then, we calculate the CEs and perform a linear fit to the CE--CE plots to obtain the CERs $E(\lambda - \GRP)/E(\GBP - \GRP)$. 
After that, we determine the base relative extinction $\ARP/E(\GBP - \GRP)$ based on classical Cepheids. 
Finally, we convert the CERs into the $A_{\lambda}/\ARP$ and determine the optical to IR extinction curves.

\begin{figure*}[ht]
\centering 
\includegraphics[angle=0,width=7.0in]{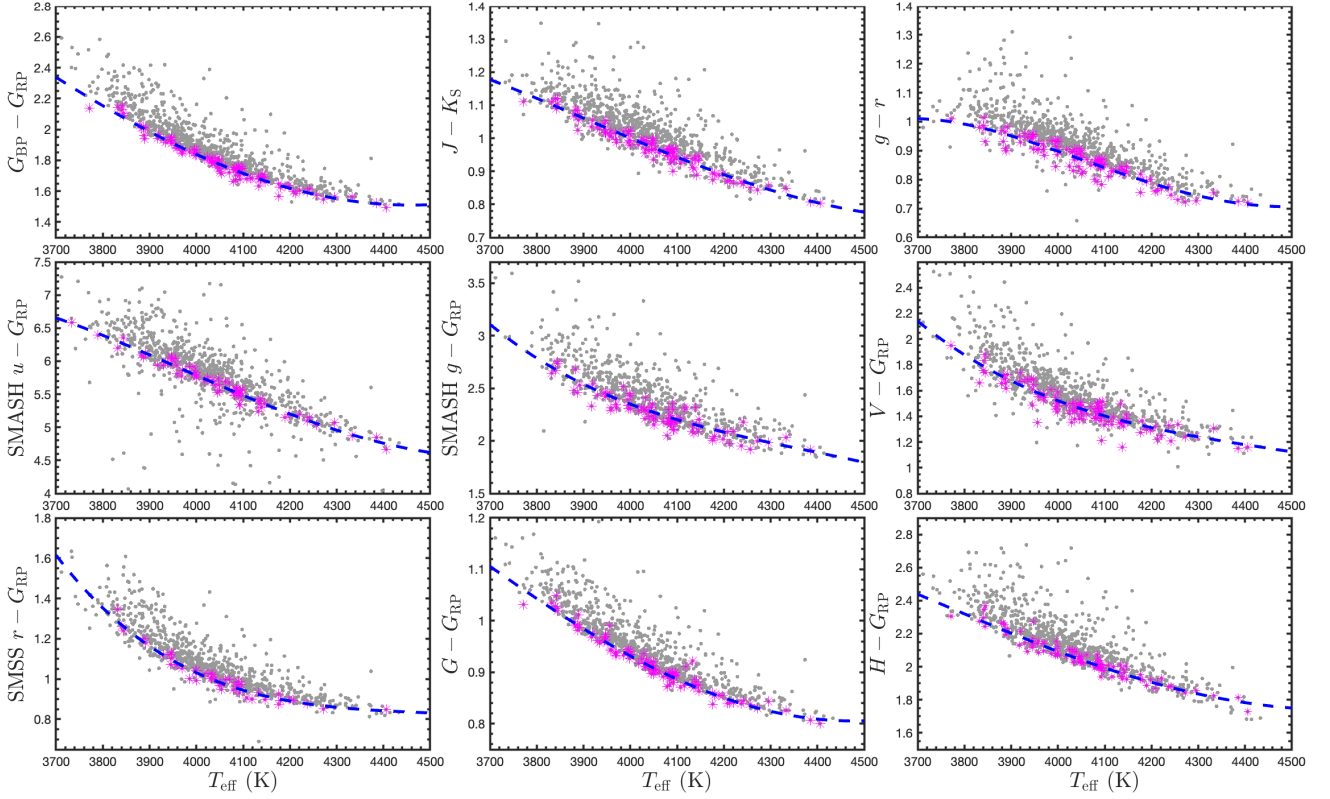} 
\vspace{-0.1in}
\caption{\footnotesize
               \label{fig:int}
$\Teff$ vs. observed color $(\lambda_1-\lambda_2)$ diagrams for the LMC RSG sample. 
Gray dots are all RSG stars. 
Magenta asterisks are the bluest stars selected according to the $\Teff$ vs. $(\GBP-\GRP)$, $(J-\Ks)$, and $(g-r)$ colors (top). 
These bluest stars are used to derive the $\Teff$--intrinsic color $(\lambda - \GRP)_0$ relations.
The blue dashed lines denote the best fit curves to these magenta asterisks with a cubic polynomial. 
}
\end{figure*}

\subsection{Intrinsic Colors}\label{intcolor}

\begin{table*}[ht]                                                                                                    
\begin{center}                                                                                                       
\caption{\label{tab:int} Coefficients of $\Teff$--$(\lambda-\GRP)_0$ Relations for RSG Stars in the LMC and SMC}
\vspace{0.1in} 
\hspace{-1.0in}                                                                                                             
\begin{tabular}{lccccccccccc}                                                                                           
\hline \hline                                                                                                        
\multicolumn{9}{c}{$(\lambda - \GRP)_0 = a_3(\log \Teff)^3+a_2(\log \Teff)^2+ a_1 (\log \Teff) + a_0$} \\
\hline                    
$(\lambda-\GRP)_0$ & $a_3$ & $a_2$ & $a_1$ & $a_0$ &RMSE  & $\Teff$ & $\Teff$& $\Teff$& $\Teff$\\ 
\hline 
\multicolumn{6}{c}{LMC} & 3800 & 4000 & 4200 & 4400 \\ 
\hline 
Gaia $\GBP$  &  635.1684    &  -6776.1429    &  24080.0209    &   -28501.8150  &0.027 & 2.154	 & 1.839	& 1.619	 & 1.516  \\     
Gaia $G$     &  263.3275    &  -2818.5802    &  10050.9654    &   -11939.5419  &0.010 & 1.043	 & 0.932	& 0.850	 & 0.809  \\     
Johnson $U$  &  -2803.2423  &  30792.3001    &  -112755.0537  &   137643.1112  &0.253 & 6.109	 & 4.973	& 4.356	 & 4.039  \\     
Johnson $B$  &  2199.0246   &  -23704.6449   &  85154.1590    &   -101937.2083 &0.063 & 3.689	 & 3.217	& 2.819	 & 2.604  \\     
Johnson $V$  &  -958.8742   &  10502.3477    &  -38348.8885   &   46684.3557   &0.059 & 1.878	 & 1.521	& 1.312	 & 1.180  \\     
SMASH $u$    &  2241.9191   &  -24231.3284   &  87271.3924    &   -104732.1293 &0.078 & 6.391	 & 5.785	& 5.203	 & 4.760  \\     
SMASH $g$    &  -1412.6610  &  15426.1534    &  -56160.0601   &   68164.7605   &0.082 & 2.786	 & 2.349	& 2.083	 & 1.890  \\     
SMSS  $v$    &  -11155.6967 &  121017.6047   &  -437611.3077  &   527497.4080  &0.349 & 5.351	 & 4.640	& 4.405	 & 3.988  \\     
SMSS  $g$    &  -0.4391     &  147.3071      &  -1058.8977    &   1925.3853    &0.041 & 2.334	 & 1.934	& 1.685	 & 1.566  \\     
SMSS  $r$    &  -1187.6736  &  13011.6972    &  -47518.0948   &   57846.6967   &0.026 & 1.353	 & 1.031	& 0.891	 & 0.842  \\     
SMSS  $i$    &  214.2653    &  -2309.2649    &  8295.6956     &   -9932.8493   &0.015 & 0.353	 & 0.345	& 0.342	 & 0.356  \\     
SMSS  $z$    &  722.5183    &  -7865.6403    &  28544.9239    &   -34532.8186  &0.012 & -0.085 & 0.032	& 0.089	 & 0.132  \\   
2MASS $J$    &  -604.4052   &  6514.7155     &  -23399.5304   &   28005.3953   &0.028 & -1.475 & -1.316	& -1.179 & -1.094 \\
2MASS $H$    &  -533.2475   &  5730.1218     &  -20514.3572   &   24466.3880   &0.032 & -2.320 & -2.093	& -1.906 & -1.783 \\
2MASS $\Ks$  &  -939.4237   &  10129.8535    &  -36397.5214   &   43575.5853   &0.030 & -2.596 & -2.316	& -2.069 & -1.900 \\
\hline                                                                                                                  
\multicolumn{6}{c}{SMC} & 4000 & 4200 & 4400 & 4600 \\                                                                                                     
\hline                                                                                                                  
Gaia $\GBP$  &  -2455.7530   &  26721.3350    &  -96926.8907   &   117205.7933  &0.035 & 1.849	& 1.651	 & 1.487	& 1.239  \\   
Gaia $G$     &  -701.8221    &  7631.5697     &  -27664.8177   &   33433.3568   &0.012 & 0.937	& 0.863	 & 0.796	& 0.703  \\   
Johnson $U$  &  -3523.5834   &  38347.8772    &  -139143.9273  &   168330.9109  &0.250 & 4.983	& 4.330	 & 3.750	& 3.072  \\   
Johnson $B$  &  -685.1667    &  7597.7117     &  -28086.9752   &   34617.1785   &0.118 & 3.228	& 2.844	 & 2.604	& 2.457  \\   
Johnson $V$  &  -347.2610    &  3886.1072     &  -14491.9800   &   18010.5421   &0.052 & 1.525	& 1.307	 & 1.192	& 1.152  \\   
SMASH $u$    &  -3883.8046   &  42302.2465    &  -153610.0378  &   185967.6808  &0.097 & 5.797	& 5.173	 & 4.654	& 4.046  \\   
SMASH $g$    &  -2981.0580   &  32438.3723    &  -117673.0032  &   142308.2294  &0.060 & 2.443	& 2.106	 & 1.817	& 1.432  \\   
SMSS  $v$    &  -25216.98533 &  274942.3707   &  -999261.2610  &   1210616.7643 &0.545 & 5.463	& 4.155	 & 3.631	& 2.622  \\   
SMSS  $g$    &  -2313.0914   &  25171.1038    &  -91313.3924   &   110432.2751  &0.056 & 1.947	& 1.714	 & 1.517	& 1.246  \\   
SMSS  $r$    &  -834.0263    &  9104.4666     &  -33133.2949   &   40199.3721   &0.041 & 1.048	& 0.918	 & 0.828	& 0.733  \\   
SMSS  $i$    &  -4.6421      &  49.6390       &  -176.8480     &   210.2539     &0.021 & 0.340	& 0.341	 & 0.342	& 0.341  \\   
SMSS  $z$    &  704.0419     &  -7677.0235    &  27905.8519    &   -33814.6040  &0.023 & 0.025	& 0.088	 & 0.128	& 0.179  \\   
2MASS $J$    &  1409.5273    &  -15354.2821   &  55756.8190    &   -67497.2622  &0.033 & -1.289	& -1.164 & -1.073	& -0.947 \\
2MASS $H$    &  2048.1671    &  -22312.5920   &  81031.0945    &   -98102.1634  &0.049 & -2.103	& -1.902 & -1.753	& -1.555 \\
2MASS $\Ks$  &  1244.4101    &  -13578.0858   &  49393.7380    &   -59906.9960  &0.040 & -2.303	& -2.061 & -1.875	& -1.679 \\
\hline                                                                                                               
\end{tabular} 
\end{center}     
\tablenotetext{}{Note: The $\Teff$--$(\lambda-\GRP)_0$ relations are applicable to estimate the intrinsic color indices of RSG stars from their $\Teff$.
For the LMC relations, they are suitable for RSG stars with parameters in the range of $3700\K \le \Teff \le 4500\K$, $0.3 < \log g < 1.0$, and $-0.6 \,\rm{dex} < [{\rm M/H}] < -0.4 \,\rm{dex}$.
For the SMC relations, they are suitable for RSG stars with parameters in the range of $3900\K \le \Teff \le 4650\K$, $0.2 < \log g < 1.0$, and $-1.0 \,\rm{dex}<[{\rm M/H}]<-0.7 \,\rm{dex}$.}                                                                                                                                                                                                      
\end{table*}

\begin{table*}[ht]                                                                                                    
\begin{center}                                                                                                       
\caption{\label{tab:int2} Other $\Teff$--Intrinsic Color Relations for RSG Stars}
\vspace{0.1in}                                                                                                           
\begin{tabular}{lcc}                                                                                           
\hline \hline
Relations & Region & Reference \\
\hline                                                                                                         
$\log \Teff=3.869-0.3360\times(V-R)_0$   &   LMC   & \citet{2003AJ....126.2867M} \\   
$\log \Teff=3.899-0.4085\times(V-R)_0$   &   SMC   & \citet{2003AJ....126.2867M} \\ 
$(J-K)_0=3.1-0.547(\Teff/1000)$   &   Milky Way   & \citet{2005ApJ...628..973L} \\ 
$\Teff=7741.9-1831.83(V-K)_0+263.135(V-K)_0^2-13.1943(V-K)_0^3$  &   Milky Way & \citet{2005ApJ...628..973L, 2006ApJ...645.1102L}\\  
$\Teff=7621.1-1737.74(V-K)_0+241.762(V-K)_0^2-11.8433(V-K)_0^3$   &   LMC & \citet{2006ApJ...645.1102L}\\
$\Teff=7167.5-1374.20(V-K)_0+157.000(V-K)_0^2-6.0481(V-K)_0^3$   &   SMC & \citet{2006ApJ...645.1102L}\\
$\Teff=8304.4-9158.6(V-R)_0+5675.2(V-R)_0^2-1194.90(V-R)_0^3$   &   Milky Way & \citet{2006ApJ...645.1102L}\\
$\Teff=7798.3-7824.4(V-R)_0+4554.8(V-R)_0^2-905.21(V-R)_0^3$   &   LMC & \citet{2006ApJ...645.1102L}\\
$\Teff=7179.4-6030.8(V-R)_0+3028.2(V-R)_0^2-525.98(V-R)_0^3$   &   SMC & \citet{2006ApJ...645.1102L}\\
$\Teff =-1746.2(J-\Ks)_0+5638.0$, valid for $0.7<(J-\Ks)_0<1.4$ mag &   LMC &\citet{2012ApJ...749..177N}\\
$\Teff =-791(J-\Ks)_0+4741$, valid for $0.8<(J-\Ks)_0<1.4$ mag   &   LMC &\citet{2019A_A...624A.128B}\\
$\Teff=-1571(J-\Ks)_0+5660$   &   SMC &\citet{2021MNRAS.502.4890D}\\
 \hline                                                                                                               
\end{tabular}                                                                                                        
\end{center} 
\tablenotetext{}{Note: These $\Teff$--intrinsic color relations are applied to estimate the $\Teff$ of RSG stars from their intrinsic color index.}                                                                                                              
\end{table*}

\citet{2001ApJ...558..309D} suggested that the stellar intrinsic color can be derived from the zero-reddening curve delineated by the blue envelope of $\Teff$ versus observed color diagrams. 
They derived the intrinsic colors for stars with all spectral types and luminosity classes in the Johnson system. 
\citet{2014ApJ...788L..12W} further developed this method and established $\Teff$--near-IR intrinsic color relations for K-type giants with $3500\K\leq\Teff\leq4800\K$.
They selected the bluest 5\% of stars in each interval of $\Delta \Teff$ = 50 K on the $\Teff$ versus observed color diagram to represent the unreddened blue edge and determined the $\Teff$--intrinsic color relation by quadratic function fitting. 
In addition, this method has been applied to derive the intrinsic colors of different types of stars in multiple bands, such as giants with $3600 \K\leq\Teff\leq5200\K$ in mid-IR bands \citep{2016ApJS..224...23X} and optical bands \citep{2017ApJ...848..106W}, dwarfs with $3850 \K\leq\Teff\leq8400 \K$ and giants with $3650 \K\leq\Teff\leq5200 \K$ in IR bands \citep{2017AJ....153....5J}, and dwarfs with $6500 \K\leq\Teff\leq8500 \K$ in UV and optical bands \citep{2018ApJ...861..153S}.

We adopted this approach to determine the intrinsic colors of our RSG stars. 
We first plotted $\Teff$ versus the observed color $(\lambda_1-\lambda_2)$ diagrams. The selection of the bluest stars can be done on any $\Teff$ vs. $(\lambda_1-\lambda_2)$ diagram. We selected the bluest 30\% of stars in each interval of $\Delta \Teff$=$50/100\K$ (LMC/SMC) on the $\Teff$ vs. 2MASS $(J-\Ks)$, Gaia $(\GBP-\GRP)$, and SMSS $(g-r)$ diagrams. Their intersection is considered to be the unreddened RSG stars, which make up about 10\% of all RSG stars. 
This process ensures that the selected bluest stars are indeed unreddened stars for a small-size sample. 
Figure~\ref{fig:int} shows $\Teff$ vs. $(\lambda_1-\lambda_2)$ diagrams for the LMC RSG sample (gray dots). The selected unreddened stars are shown as magenta asterisks (top panels of Figure~\ref{fig:int}). 
These stars are also located at the edge of the bluest colors in other bands (see the middle and bottom panels of Figure~\ref{fig:int}). We fitted these stars with a cubic polynomial function to obtain the $\Teff$--$(\lambda - \GRP)_0$ relations (see blue dashed lines in Figure~\ref{fig:int}). 
Finally, the intrinsic color indices $(\lambda - \GRP)_0$ are derived for three Johnson bands $U$, $B$, $V$, 
two Gaia bands $\GBP$, $G$, 
two SMASH bands $u$, $g$, 
five SMSS bands $v$, $g$, $r$, $i$, $z$,  
and three 2MASS bands $J, H$, $\Ks$. 
The intrinsic color indices can be expressed as:  
\begin{eqnarray} \label{equ_int}
(\lambda - \GRP)_0  & = &
       a_3(\log \Teff)^3+a_2(\log \Teff)^2\nonumber\\ 
& &  + a_1 (\log \Teff) + a_0~~,
\end{eqnarray}
where the corresponding coefficients for each color index are listed in Table~\ref{tab:int}. The root-mean-square errors (RMSEs) of these equations are also list in this table. 
Other reported $\Teff$--intrinsic color relations for RSG stars are presented in Table~\ref{tab:int2}. The corresponding applicable regions and references are also listed.

Compared to the previous $\Teff$--intrinsic color relations, our relations are based on a larger sample and takes into account higher-order terms. 
According to the common bands, we compared the relations of $\Teff$ vs. $(J-\Ks)_0$ and $\Teff$ vs. $(V-\Ks)_0$. 
For $(J-\Ks)_0$, our SMC result is in excellent agreement ($\Delta=0.017\pm0.020$ mag) with that of \citet{2021MNRAS.502.4890D}, and our LMC result is consistent ($\Delta=-0.066\pm0.004$ mag) with that of \citet{2012ApJ...749..177N}. 
The overall trend of \citet{2019A_A...624A.128B}'s result is different from the literature's results and ours (see Fig. 3 of his paper for details). 
For $(V-\Ks)_0$, the overall trend of our LMC and SMC results are consistent with \citet{2006ApJ...645.1102L}, with an overall deviation of $\sim0.3$ mag ($\Delta=-0.35\pm0.07$, $\Delta=-0.23\pm0.02$). 
This systemic bias in intrinsic color indices does not affect the further determination of the CER, which is the slope of a linearly fit of the CE--CE plot. The scatters are comparable to the RMSEs of the relations.

In the optical and $u$ band, the intrinsic color index depends not only on the $\Teff$ but also on the [M/H] and $\log g$. 
We examined the effects of [M/H] and $\log g$. We found that their effects on the derivation of the intrinsic colors are negligible due to the very narrow range of [M/H] and $\log g$ values of the RSG stars we chose.

Another advantage of our intrinsic color calculation method is that we no longer need to subtract the Galactic extinction separately. 
The observed magnitude $m_\lambda$ of RSG stars in the LMC/SMC is composed of three components: the absolute magnitude $m_0$, the Galactic extinction $A_{\rm MW}$, and the local extinction $A_{\rm MC}$. 
In this section, we establish the $\Teff$--intrinsic color relations based on the selected unreddened RSG stars in the $\Teff$ vs. $(\lambda_1-\lambda_2)$ diagram. 
These selected unreddened sources do not suffer from the local extinction, namely $A_{\rm MC}=0$.  However, the foreground Galactic extinction $A_{\rm MW}$ of these stars is not zero. 
Therefore, the $\Teff$--intrinsic color relations also contain the effects of the foreground Galactic extinction. 
In subsequent CE calculations, the CE is the observed color index (including $A_{\rm MW}$ and $A_{\rm MC}$) minus the intrinsic color index (including $A_{\rm MW}$). 
The resulting CE naturally subtracts the foreground Galactic extinction. 
Therefore, we no longer need to assume the foreground extinction of the Milky Way and avoid introducing additional uncertainties in the CE calculation.

\subsection{Color Excess Ratios}\label{CEratio}

\begin{figure*}[ht]
\centering
\vspace{-0.0in}
\includegraphics[angle=0,width=6.8in]{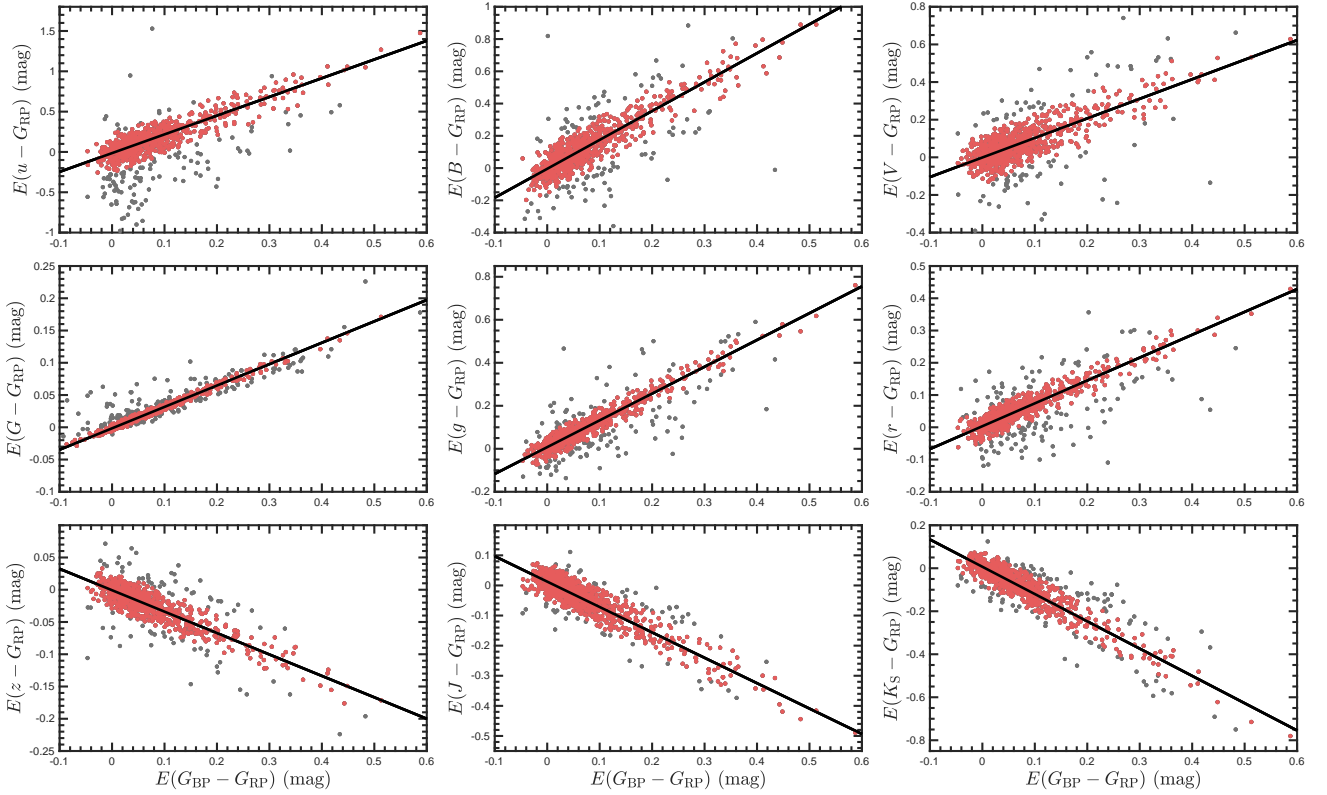}
\vspace{-0.0in}
\caption{\footnotesize
               \label{fig:EE1}
The color excess--color excess diagrams  $E(\GBP - \GRP)$ vs. $E(\lambda - \GRP)$ of RSG stars in the LMC,  
where $\lambda$ are 
$u$ from SMASH, $B$ and $V$ from MCPS, $G$ from Gaia, 
$g$, $r$, and $z$ from SMSS, 
and $J$ and $\Ks$ from 2MASS, respectively, from the top left to the bottom right. 
The gray dots are stars dropped based on a $2\sigma$ criterion. 
The black lines are the best fits to the data (red dots), and the slopes are CERs listed in Table~\ref{tab:ext}.}
\end{figure*}

Based on the $\Teff$--$(\lambda-\GRP)_0$ relations, the $(\lambda-\GRP)_0$ is derived. 
The observed color $(\lambda-\GRP)$ is the difference between two observed magnitudes. 
The CE is then determined by 
$E(\lambda-\GRP)=(\lambda-\GRP)-(\lambda-\GRP)_0$. 
We performed a linear fit of the CE--CE plots to obtain the CERs.  
Figure~\ref{fig:EE1} is an example that displays the CER $E(\lambda-\GRP)/E(\GBP - \GRP)$ determination for RSG stars in the LMC, where $\lambda$ are 
SMASH $u$, Johnson $B$, $V$, Gaia $G$, SMSS $g$, $r$, $z$,   
and 2MASS $J$, $\Ks$ bands, respectively, from the top left to the bottom right. 
The CE--CE diagrams of LMC RSG stars in the other bands and SMC RSG stars are presented in Appendix Figures~\ref{fig:EE2}, ~\ref{fig:EE3}, and~\ref{fig:EE4}.  
The gray dots are the objects with deviations exceeding $2\sigma$, which are excluded from a statistical point of view.
The red dots are the final objects used to perform the linear fit, while the black lines are the best linear fit lines. 
The CERs are the slopes of these lines.

Since many RSG stars in MCs are located in the low-extinction region, and the number of stars with different extinction is not uniformly distributed, the formal error of the CER $\sigma_1$ obtained by fitting all RSG stars may be underestimated relative to the actual error. We subdivided the samples by $(\GBP - \GRP)>-0.05+0.01*n$, where $n$ varies from 0 to 29 for LMC, from 0 to 14 for SMC. A linear fit was applied to each subsample to determine the slope. To avoid the effect on the slope due to the variation of the intercept, we fix the intercept. This fixed intercept value was determined by the whole sample and was very close to 0. The standard deviation of the slopes $\sigma_2$ was determined. The errors in photometry, as well as the errors in estimating the intrinsic color $(\lambda-\GRP)_0$, are also propagated to the slope. the errors in photometry for the Gaia and SMSS bands are small, ranging from 0.003 to 0.02 mag, and for the 2MASS bands, ranging from 0.02 to 0.03 mag. The mean error in the $U$ band is the largest, around 0.07 mag. Among the intrinsic colors, errors of $(G-\GRP)_0$ and $(\GBP-\GRP)_0$ are the smallest at 0.01 to 0.02 mag. The intrinsic color errors in the UV bands are the largest, reaching 0.2 to 0.3 mag. We used Monte Carlo simulation to estimate the propagated errors. 
In the simulation, the number of data points and the amount of extinction were kept consistent with the observations, and the error on the $x,y$-axes is the larger one among the photometric error and the intrinsic color error. 
We performed a linear fit to the simulated data points to obtain the slope. Then we repeated the simulation 1000 times and calculate the standard deviation of the slope $\sigma_3$. This standard deviation was found to be comparable or slightly smaller than the observation-based slope error $\sigma_2$. This is understandable because when we choose the CE $E(\GBP - \GRP)$ with the smallest error as the $x$-axis, the error in the $y$-axis will be fully considered in the fit. The relevant analysis can also be found in Section 4.2 of \citet{2019ApJ...877..116W}. The final error of the CER is estimated by $\sigma_{\rm CER}=\sqrt{\sigma_1^2+\max(\sigma_2,\sigma_3)^2}$. For each band, the determined CERs and their errors are listed in Table~\ref{tab:ext}.

\begin{figure}[ht]
\centering
\vspace{-0.0in}
\includegraphics[angle=0,width=3.2in]{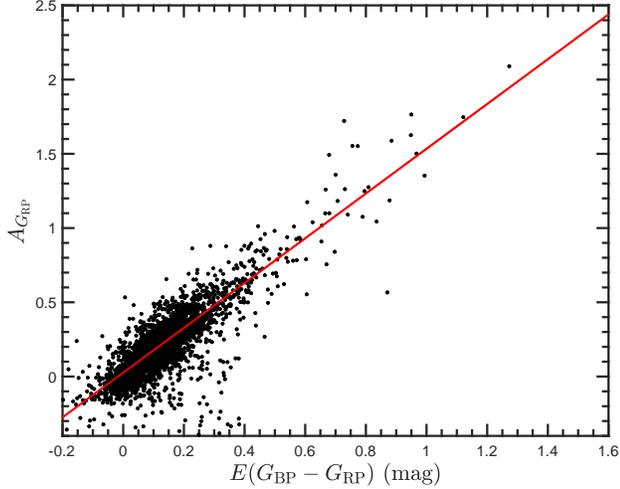}
\vspace{-0.0in}
\caption{\footnotesize
               \label{fig:cep_ext}
The color excess--extinction diagram $E(\GBP - \GRP)$ vs. $\ARP$ of 3181 Cepheids in the LMC.  
The red line is a linear fit, which denotes the extinction direction. 
}
\end{figure}

\begin{table*}[ht]
\begin{center}
\caption{\label{tab:ext} Multi-band Color Excess Ratios and Extinction Coefficients}
\vspace{0.1in} 
\hspace{-1.0in}    
\begin{tabular}{lccccc}
\hline \hline                                                      
Band ($\lambda$) & $\lambda_{\rm eff, 0}$ ($\mum$) & $E(\lambda-\GRP)/E(\GBP-\GRP)$ &  $A_\lambda/\ARP$ &  $A_\lambda/A_V$  &  $A_\lambda/E(\GBP-\GRP)$ \\ 
\hline 
\multicolumn{6}{c}{LMC}  \\                                                                                                     
\hline   
 Gaia $\GBP$       &  0.5588&  ...               &  $1.629 \pm0.015$  &  $0.985\pm0.009$  &  $2.589\pm0.014$  \\  
 Gaia $\GRP$       &  0.7824&  ...               &         1          &  $0.605\pm0.005$  &  $1.589\pm0.014$  \\  
 Gaia $G$          &  0.6816&  $ 0.332\pm0.007$  &  $1.209 \pm0.011$  &  $0.731\pm0.007$  &  $1.921\pm0.016$  \\  
Johnson $U$        &  0.3729&  $ 2.333\pm0.119$  &  $2.468 \pm0.078$  &  $1.492\pm0.046$  &  $3.922\pm0.120$  \\  
Johnson $B$        &  0.4599&  $ 1.781\pm0.032$  &  $2.121 \pm0.027$  &  $1.282\pm0.016$  &  $3.370\pm0.035$  \\  
Johnson $V$        &  0.5572&  $ 1.040\pm0.035$  &  $1.654 \pm0.026$  &  $      1      $  &  $2.629\pm0.037$  \\  
SMASH $u$          &  0.3836&  $ 2.323\pm0.056$  &  $2.462 \pm0.041$  &  $1.488\pm0.024$  &  $3.912\pm0.058$  \\  
SMASH $g$          &  0.4964&  $ 1.403\pm0.037$  &  $1.883 \pm0.028$  &  $1.138\pm0.017$  &  $2.992\pm0.039$  \\  
SMSS $v$           &  0.3875&  $ 2.224\pm0.262$  &  $2.400 \pm0.166$  &  $1.451\pm0.099$  &  $3.813\pm0.262$  \\  
SMSS $g$           &  0.5249&  $ 1.246\pm0.014$  &  $1.784 \pm0.018$  &  $1.078\pm0.011$  &  $2.835\pm0.020$  \\  
SMSS $r$           &  0.6204&  $ 0.704\pm0.011$  &  $1.443 \pm0.014$  &  $0.873\pm0.009$  &  $2.293\pm0.018$  \\  
SMSS $i$           &  0.7783&  $ 0.045\pm0.007$  &  $1.028 \pm0.010$  &  $0.622\pm0.006$  &  $1.634\pm0.016$  \\  
SMSS $z$           &  0.9135&  $-0.329\pm0.006$  &  $0.793 \pm0.008$  &  $0.479\pm0.005$  &  $1.260\pm0.015$  \\  
2MASS $J$          &  1.2345&  $-0.863\pm0.018$  &  $0.457 \pm0.012$  &  $0.276\pm0.007$  &  $0.726\pm0.023$  \\  
2MASS $H$          &  1.6393&  $-1.150\pm0.029$  &  $0.276 \pm0.018$  &  $0.167\pm0.011$  &  $0.439\pm0.032$  \\  
2MASS $\Ks$        &  2.1757&  $-1.339\pm0.028$  &  $0.157 \pm0.018$  &  $0.095\pm0.010$  &  $0.250\pm0.031$  \\  
\hline                                                                                                             
\multicolumn{6}{c}{SMC}  \\                                                                                                     
\hline   
 Gaia $\GBP$       &  0.5527&  ...               &  $1.708\pm0.052$  &  $1.006\pm0.029$  &  $2.412\pm0.041$  \\  
 Gaia $\GRP$       &  0.7753&  ...               &        1          &  $0.589\pm0.014$  &  $1.412\pm0.041$  \\  
 Gaia $G$          &  0.6668&  $ 0.346\pm0.008$  &  $1.245\pm0.030$  &  $0.734\pm0.018$  &  $1.758\pm0.042$  \\  
Johnson $U$        &  0.3719&  $ 2.238\pm0.467$  &  $2.585\pm0.336$  &  $1.522\pm0.195$  &  $3.650\pm0.469$  \\  
Johnson $B$        &  0.4568&  $ 1.810\pm0.102$  &  $2.282\pm0.090$  &  $1.344\pm0.053$  &  $3.222\pm0.110$  \\  
Johnson $V$        &  0.5552&  $ 0.985\pm0.104$  &  $1.698\pm0.084$  &  $      1      $  &  $2.397\pm0.112$  \\  
SMASH $u$          &  0.3834&  $ 3.006\pm0.193$  &  $3.129\pm0.156$  &  $1.843\pm0.091$  &  $4.418\pm0.198$  \\  
SMASH $g$          &  0.4946&  $ 1.367\pm0.182$  &  $1.968\pm0.137$  &  $1.159\pm0.080$  &  $2.779\pm0.187$  \\  
SMSS $v$           &  0.3866&  $ 2.581\pm0.309$  &  $2.828\pm0.229$  &  $1.666\pm0.133$  &  $3.993\pm0.312$  \\  
SMSS $g$           &  0.5222&  $ 1.251\pm0.046$  &  $1.886\pm0.056$  &  $1.111\pm0.032$  &  $2.663\pm0.062$  \\  
SMSS $r$           &  0.6189&  $ 0.646\pm0.033$  &  $1.457\pm0.042$  &  $0.858\pm0.024$  &  $2.058\pm0.053$  \\  
SMSS $i$           &  0.7761&  $-0.014\pm0.028$  &  $0.990\pm0.031$  &  $0.583\pm0.018$  &  $1.398\pm0.050$  \\  
SMSS $z$           &  0.9129&  $-0.391\pm0.029$  &  $0.723\pm0.027$  &  $0.426\pm0.015$  &  $1.021\pm0.050$  \\  
2MASS $J$          &  1.2345&  $-0.851\pm0.058$  &  $0.397\pm0.042$  &  $0.234\pm0.024$  &  $0.561\pm0.071$  \\  
2MASS $H$          &  1.6393&  $-1.041\pm0.086$  &  $0.263\pm0.061$  &  $0.155\pm0.036$  &  $0.371\pm0.095$  \\  
2MASS $\Ks$        &  2.1757&  $-1.216\pm0.092$  &  $0.139\pm0.065$  &  $0.082\pm0.038$  &  $0.196\pm0.101$  \\ 
\hline  
\end{tabular}
\end{center}
\end{table*}

\subsection{Relative Extinction}\label{A/A}

The relative extinction $A_{\lambda}/\ARP$ can be converted from the CER by the equation 
\begin{equation} \label{equ2}
\frac{A_{\lambda}}{\ARP} = 1 + k_\lambda \frac{E(\GBP - \GRP)}{\ARP}~~,
\end{equation} 
where $k_\lambda$ represents the CER $\frac{E(\lambda - \GRP)}{E(\GBP - \GRP)}$. 
To obtain the value of $\frac{A_{\lambda}}{\ARP}$, $\frac{\ARP}{E(\GBP - \GRP)}$ is required. 
The relative extinction $\frac{\ARP}{E(\GBP - \GRP)}$ or $\frac{\ABP}{\ARP}$ is the base extinction value. 

The accuracy of the relative extinction is determined by the accuracy of the stellar distance and absolute magnitude estimations. For the LMC, the distance dispersion of young stars is small, while for the SMC, the distance dispersion will be larger. For the absolute magnitudes, the error of the best absolute magnitudes obtained from the spectra is about 0.2 to 0.3 mag. Considering the small extinction of LMC and SMC, this error implies that we will not be accurate enough to estimate the relative extinction using spectroscopic RSG stars. For this reason, we consider using the classical Cepheid to estimate the relative extinction $\frac{\ARP}{E(\GBP - \GRP)}$, since the period--luminosity relation of Cepheids can predict an absolute magnitude with an error of only 0.08 mag.

We used Cepheids in Gaia DR3 \citep{2022arXiv220606212R} and established period--color relations $\GBP-\GRP=f(\log P)$ for the fundamental mode and the first-overtone mode Cepheids, respectively. Then Cepheids with low extinction were selected by $-0.16<(\GBP-\GRP)-f(\log P)<-0.08$, where 0.08 is the $1\sigma$ dispersion of the period--color relations. The selected low-extinction Cepheids account for about 15\% of the total Cepheids, and they were used to determine the period--intrinsic color relation $\log P$--$(\GBP-\GRP)_0$ and the period--luminosity relation $\log P$--$\MRP$. These relations were then used to estimate $E(\GBP-\GRP)$ and $\ARP$ for all Cepheids. 
For example, our determined period--intrinsic color relation for LMC fundamental Cepheids is $(\GBP-\GRP)_0=(0.263\pm0.006)\log P+(0.606\pm0.004)$. Please note that this relation may deviate from the optimal $\log P$--$(\GBP-\GRP)_0$ relation by a small amount, so it is only suitable for analyzing extinction laws and not for measuring the absolute amounts of extinction. 
Another method to obtain the $\log P$--$(\GBP-\GRP)_0$ and $\log P$--$\MRP$  relations is subtracting the extinction obtained from the external extinction map. 
We derived $(\GBP-\GRP)_0=(0.278\pm0.008)\log P+(0.575\pm0.005)$, which is close to the relation determined by the former method. 
The difference between these two methods in determining the value of $\frac{\ARP}{E(\GBP - \GRP)}$ is very small. 
This is because $\frac{\ARP}{E(\GBP - \GRP)}$ is the slope of two sets of extinction (i.e., Figure~\ref{fig:cep_ext}), which is largely independent of the zero points of the intrinsic color and absolute magnitude.

We performed a linear fit to 3181 LMC Cepheids (fundamental mode and first-overtone mode Cepheids) and 1906 SMC Cepheids (only fundamental mode Cepheids) with positive extinction and CE values. Figure~\ref{fig:cep_ext} shows an example of determining $\frac{\ARP}{E(\GBP - \GRP)}$. The results for LMC and SMC are $\ARP=(1.508\pm0.013) \times E(\GBP-\GRP)+0.026, \sigma_{\rm LMC}=0.082$ and $\ARP=(1.340\pm0.041) \times E(\GBP-\GRP)+0.068, \sigma_{\rm SMC}=0.147$, respectively. $\sigma_{\rm LMC}=0.082$ is dominated by the error in the estimation of the Cepheid's absolute magnitude, while $\sigma_{\rm SMC}=0.147$ contains an additional distance modulus dispersion of about 0.12 mag for the SMC Cepheids. Nevertheless, they are both much smaller than the errors in the RSG absolute magnitudes estimated with spectral parameters ($0.2-0.3$ mag). If we used the relations $\log P$--$(\GBP-\GRP)_0$ and $\log P$--$\MRP$ determined by the external extinction map, the extinction laws obtained are $\ARP=(1.498\pm0.013) \times E(\GBP-\GRP)$ and $\ARP=(1.286\pm0.041) \times E(\GBP-\GRP)$ for LMC and SMC. The difference is $0.6\sigma$ and $0.9\sigma$, which indicates that the extinction law is little affected by the zero point of the intrinsic colors and absolute magnitudes.

The relative extinction $\ARP/E(\GBP-\GRP)$ based on Cepheids needs a small correction to be used for RSG stars, since the temperature of Cepheids is about 1500 K higher. We adopted the synthetic stellar spectra \citep{1997A&AS..125..229L} $F_\lambda$ for the Cepheid and the RSG with $T_{\rm eff}$, $\log g$, and [Fe/H] values according to the average stellar parameters.  For LMC Cepheids, we adopted $T_{\rm eff}=5500$ K, $\log g=1.5$ and [Fe/H] $= -0.5$, while for LMC RSG stars, we adopted $T_{\rm eff}=4000$ K, $\log g=0.5$ and [Fe/H] $= -0.5$. For SMC Cepheids, we adopted $T_{\rm eff}=5500$ K, $\log g=1.5$ and [Fe/H] $= -1.0$, while for SMC RSG stars, we adopted $T_{\rm eff}=4250$ K, $\log g=0.5$ and [Fe/H] $= -1.0$. Based on these synthetic stellar spectra, an $\Rv=3.1$ extinction law was adopted to simulate the extinction $\ARP$ and the CE $E(\GBP-\GRP)$. The correction was determined by comparing the two simulated $\frac{\ARP}{E(\GBP - \GRP)}$ values of Cepheids and RSG stars. The choice of different $\Rv$ extinction curves has little effect on this correction. After correction, the relative extinctions are $\frac{\ARP}{E(\GBP - \GRP)}=(1.589\pm0.014)$ and $\frac{\ARP}{E(\GBP - \GRP)}=(1.412\pm0.041)$ for the LMC and SMC RSG stars. Here, we took half of the grid point interval ($125$ K) as the error in temperature and calculated the propagating errors (0.004 and 0.005 in $\frac{\ARP}{E(\GBP - \GRP)}$ for LMC and SMC) to the correction.

Combining the determined $\ARP/E(\GBP-\GRP)$ with the CERs derived in Section~\ref{CEratio}, we determined the optical to IR multi-band relative extinction $A_{\lambda}/\ARP$ by Equation~\ref{equ2}. 
We uniformly calculated the static effective wavelength $\lambda_{\rm eff, 0}$ of RSG stars through 
\begin{equation} \label{equ3}
\lambda_{\rm eff, 0}=\frac{\int \lambda F_\lambda(\lambda)S(\lambda)d\lambda} 
{\int F_\lambda(\lambda)S(\lambda)d\lambda}~~.  
\end{equation}
$F_\lambda(\lambda)$ is the synthetic stellar spectra according to the average stellar parameters of RSG stars, and $S(\lambda)$ is the filter transmission curve. 
Finally, the effective wavelengths $\lambda_{\rm eff, 0}$, CERs, relative extinction $A_{\lambda}/\ARP$, $A_{\lambda}/\Av$, and extinction coefficients $A_{\lambda}/E(\GBP-\GRP)$ are tabulated in Table~\ref{tab:ext}.

\section{Results}\label{extlaw} 
\subsection{$\Rv$-dependent Optical--IR Extinction}\label{Rvlaw}

The Galactic wavelength-dependent extinction curves can be described by a one-parameter function of $\Rv$. 
$\Rv=\AV/E(B-V)=\AV/(\AB-\AV)$ is the ratio of the total extinction to the selective extinction \citep[][hereafter CCM]{1989ApJ...345..245C}. 
The average extinction curve for the Galactic diffuse interstellar medium (ISM) is $\Rv \sim 3.1$ \citep{1989ApJ...345..245C, 2007ApJ...663..320F,2011ApJ...737..103S}. 
\citet{2019ApJ...877..116W} investigated the Galactic extinction law and obtained an average extinction curve with significantly improved accuracy. 
The determined extinction curve can be expressed as an $\RV=3.16\pm0.15$ curve, but is at least $10-20\%$ lower than the CCM extinction curve at wavelengths longer than 600 nm.

\begin{figure*}[ht]
\centering
\vspace{0.1in} 
\includegraphics[angle=0,width=4.5in]{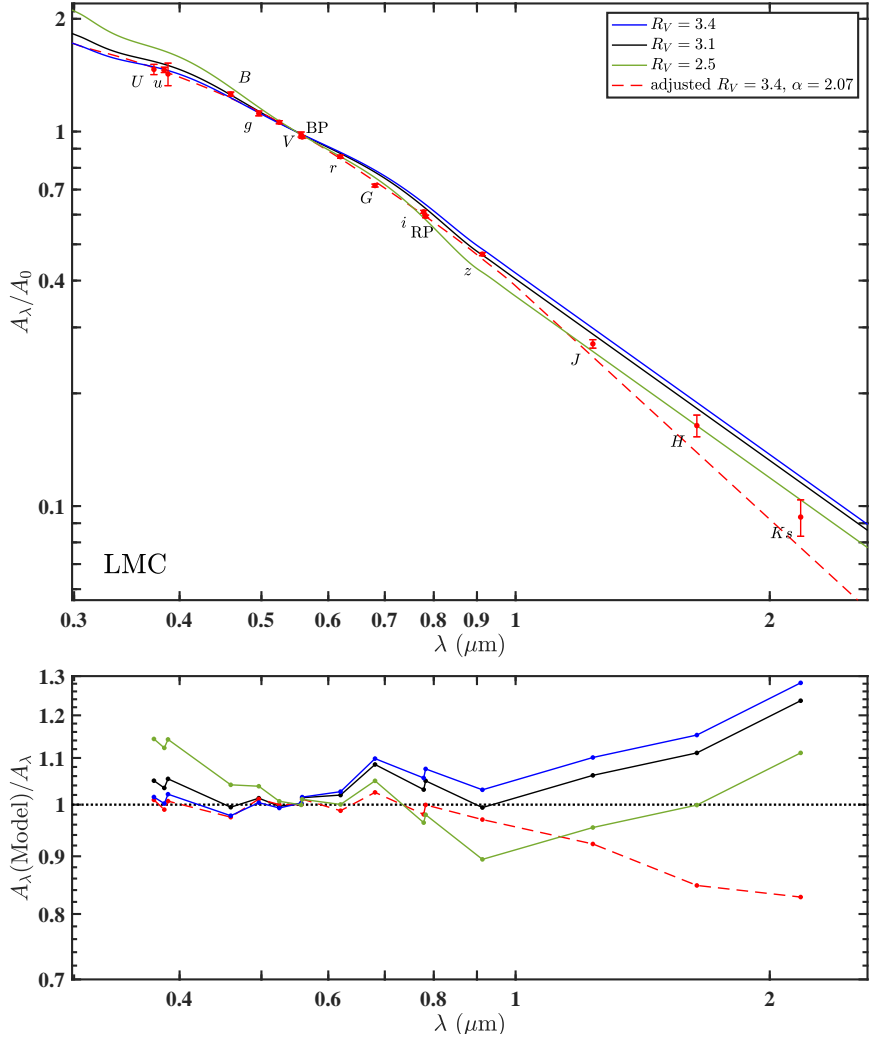}
\vspace{-0.0in}
\caption{\footnotesize
               \label{fig:ext} 
The optical--IR extinction law of the LMC. 
Top panel: our determined multi-band relative extinctions $A_{\lambda}/A_0$ (red filled circles with red error bars). 
$A_0$ is the extinction at wavelength 550 nm with negligible bandwidth.
For comparison, the CCM $\Rv$ = 3.4 (blue line), 3.1 (black line), 2.5 (green line) model extinction curves are also shown. 
The observed relative extinctions at $\lambda\le0.9\mum$ can be best fitted by the adjusted $\Rv=3.4$ curve  (red dashed line).  
Bottom panel: comparison of model extinction curves to the observed extinction in ratio.
}
\end{figure*}

\begin{figure*}[ht]
\centering
\vspace{0.1in} 
\includegraphics[angle=0,width=4.5in]{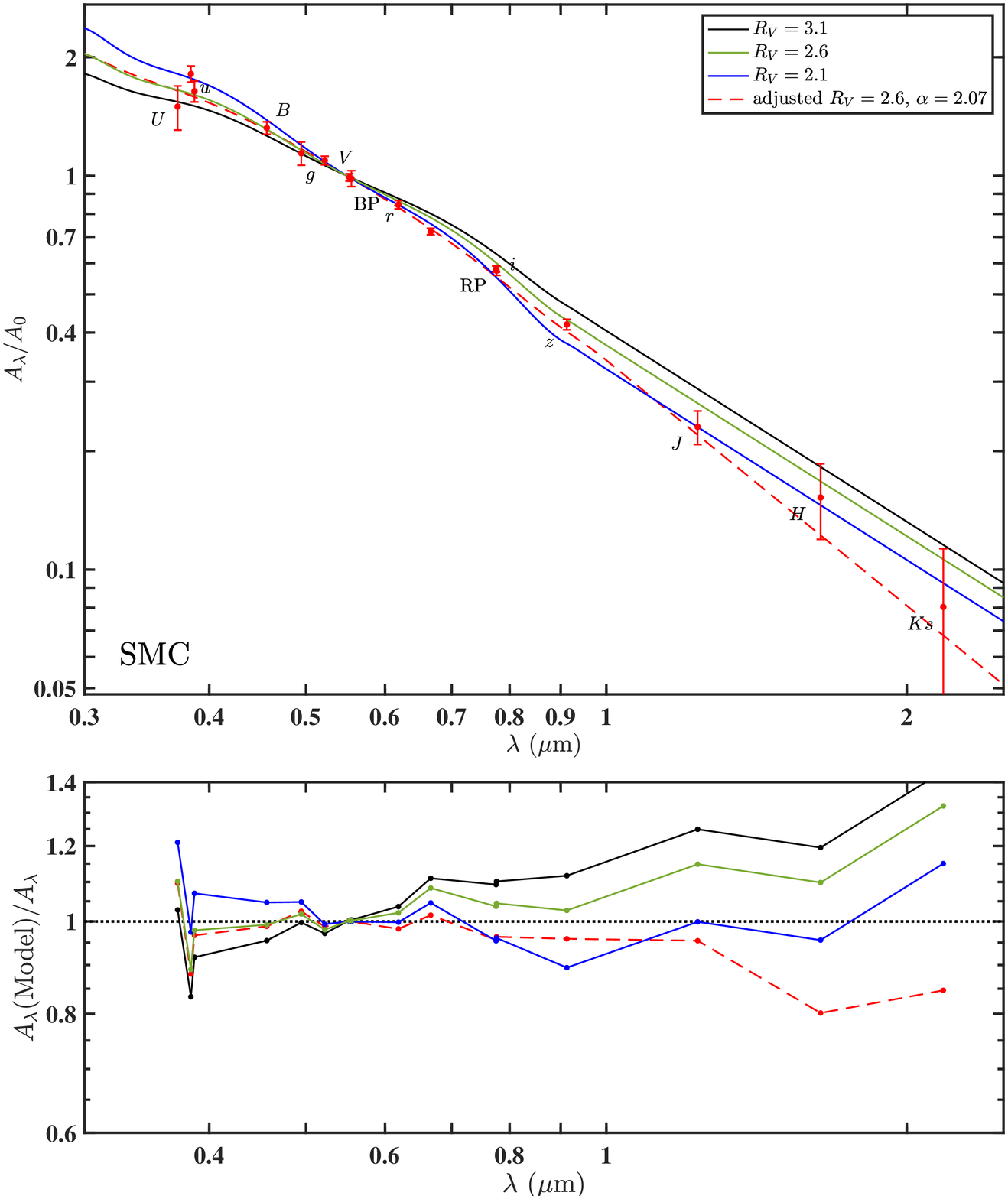}
\vspace{-0.0in}
\caption{\footnotesize
               \label{fig:ext_smc} 
The optical--IR extinction law of the SMC. 
Top panel: our determined multi-band relative extinctions $A_{\lambda}/A_0$ (red filled circles with red error bars). 
$A_0$ is the extinction at wavelength 550 nm with negligible bandwidth.
For comparison, the CCM $\Rv$ = 3.1 (black line), 2.6 (green line), and 2.1 (blue line) model extinction curves are also shown. 
The observed relative extinctions at $\lambda\le0.9\mum$ can be best fitted by the adjusted $\Rv=2.6$ curve  (red dashed line). 
Bottom panel: comparison of model extinction curves to the observed extinction in ratio.
}
\end{figure*}

To determine the $\Rv$-dependent extinction law of the LMC and SMC, we compared our multi-band extinction coefficients with different CCM model extinction curves shown in Figure~\ref{fig:ext} and Figure~\ref{fig:ext_smc}. 
Our derived optical to IR extinction coefficients are shown as red-filled circles with red error bars. 
For LMC, the CCM $\Rv$ = 3.4, 3.1, and 2.5 model extinction curves are plotted as blue, black, and green lines, respectively.  
The larger the $\RV$ value, the flatter the curve.
The bottom panel of Figure~\ref{fig:ext} presents the ratio of the model extinction coefficient from CCM $\Rv$ = 3.4 (blue), 3.1 (black), and 2.5 (green) curves to our observed extinction coefficient $A_\lambda$ (Model)/$A_\lambda$ in each band. 
Figure~\ref{fig:ext_smc} is for SMC, and the CCM $\Rv$ = 3.1, 2.6, and 2.1 model extinction curves are plotted as black, green, and blue lines, respectively.  
The bottom panel of Figure~\ref{fig:ext_smc} shows the comparison of the model extinction coefficient from CCM $\Rv$ = 3.1 (black), 2.6 (green), and 2.1 (blue) curves with our observed results $A_\lambda$ (Model)/$A_\lambda$.

In the wavelength range of $300-550$ nm, the extinction law of both LMC and SMC deviates slightly from the CCM $\RV=3.1$ curve. The LMC extinction law is close to the CCM $\RV=3.4$ curve, while the LMC extinction law is close to the $\RV=2.6$ curve.

In the $550-900$ nm wavelength range, our LMC extinction coefficients are up to $10\%$ lower than the CCM $\RV=3.4$ extinction coefficients (Figure~\ref{fig:ext}). 
The SMC extinction curve shows similar results, i.e., the observed extinction is $10\%$ lower than the CCM $\RV=2.6$ extinction curve (Figure~\ref{fig:ext_smc}).
These imply that the CCM extinction curves need to be adjusted in the wavelength range of $550-900$ nm to match observations. This adjustment is also been suggested in the analysis of the extinction curve of the Milky Way \citep{2019ApJ...877..116W}. We speculate that the CCM extinction law is slightly overestimated in this wavelength range since their result is based on only one band ($R$ band, 770 nm).

In IR bands ($0.9\mum < \lambda < 3\mum$), the extinction law follows a power law $A_{\lambda}\propto{\lambda^{-\alpha}}$, independent of the adopted $\Rv$ value.  
The power-law index of the CCM extinction curve is a fixed value of $\alpha=1.61$. 
Based on accurate photometry and parallax of Gaia, a pure RC sample from APOGEE, and a robust determination method, \citet{2019ApJ...877..116W} recommended a steep average near-IR extinction with $\alpha=2.07$. 
Compared to the value of $\alpha=1.61$, recent studies reported larger values, ranging from 1.9 to 2.6 \citep[see][and references therein]{2018ApJ...859..137C,2019ApJ...877..116W,2021ApJ...906...73H,2022MNRAS.514.2407S}. 
The uncertainties of our IR measurements are large. It is because the IR extinction and CEs of LMC and SMC are generally small, resulting in a large uncertainty in the CER measurements. 
Therefore, in this work, we did not attempt to constrain the power-law index of the near-IR extinction law. 
It is worth noting that for the IR extinction correction of the MCs, if the $V$-band extinction is $\AV$=1 mag, then the difference in the $\Ks$-band extinction obtained using different extinction laws (with a power-law index of $\alpha$ = 1.61 or 2.07) is only 0.02 mag, which is comparable to the photometric error.

The mid-IR extinction in $3\mum < \lambda < 30\mum$ does not approach zero with increasing wavelength. 
Many observations show that the Galactic extinction law in $3\mum < \lambda < 8\mum$ is relatively flat across a diversity of sight lines and values of $\RV$ \citep[see][and references therein]{2009ApJ...707...89G, 2013ApJ...773...30W, 2015ApJ...811...38W, 2016ApJS..224...23X, 2021ApJ...906...73H}. 
For the LMC extinction, \citet{2013ApJ...776....7G} also reported a flat $3-8\mum$ extinction law. 
At long wavelengths ($8-30\mum$), the Galactic extinction is dominated by two silicate absorption features centered at 9.7$\mum$ and 18$\mum$ \citep{2003ARA&A..41..241D}.

CCM extinction law contains two main functions: A and B. Function A is the fundamental $\Rv=3.1$ extinction curve, while function B shows the offset of any given $\Rv$ from the $\Rv=3.1$ extinction curve.
To better describe the observed extinction, we adjusted the parameters of CCM extinction law . 
For function A, we kept it the same as \citet[][see Equation 9 of their work]{2019ApJ...877..116W}, which has been adjusted according to the Galactic extinction law. For function B, we kept it the same as the CCM in the optical band, while adjusted it in the near-IR band to ensure a smooth extinction profile. The final adjusted $\RV$-dependent extinction laws are shown below. 
\begin{eqnarray} \label{equ4}
 A_{\lambda}/A_{\rm V}=A+B/R_{\rm V} ~~; 
\end{eqnarray} 
Optical: $0.3\,{\rm \mu m} <\lambda< 1.0\,{\rm \mu m}$ and $Y=1/\lambda(\,{\rm \mu m})-1.82$, 
\begin{eqnarray} \label{equ5}
A  & = & 1.0+0.7499Y-0.1086Y^2-0.08909Y^3 \nonumber\\
                             & &  +0.02905Y^4+0.01069Y^5 \nonumber\\
                             & &  +0.001707Y^6-0.001002Y^7 ~~; 
\end{eqnarray}
\begin{eqnarray} \label{equ6}
 B & = & (1.41338Y+2.28305Y^2+1.07233Y^3 \nonumber\\ 
   & &  -5.38434Y^4-0.62251Y^5+5.30260Y^6 \nonumber\\ 
   & & -2.09002Y^7)\times(1-R_{\rm V}/3.1)~~. 
\end{eqnarray} 
Near-IR: $1.0\,{\rm \mu m} \leq \lambda< 3.33\,{\rm \mu m}$,
\begin{equation} \label{equ7}
A  = (0.3722\pm0.0026) \lambda^{-2.07\pm0.03}~~; 
\end{equation}
\begin{equation} \label{equ8}
B = (-0.5182\pm0.0067)\lambda^{-2.07\pm0.03}\times(1-R_{\rm V}/3.1)~~. 
\end{equation} 
As seen in Figure~\ref{fig:ext} and Figure~\ref{fig:ext_smc}, our adjusted $\Rv=3.4$ and $\Rv=2.6$ extinction curves (red dashed lines) are more consistent with the observations than the corresponding CCM extinction curves. Compared to observations, the maximum inconsistency of our extinction curves in the optical band is only $3\%$ for LMC.

We estimated the $\Rv$ values of LMC and SMC by comparing the observed relative extinction with the adjusted $\Rv$-dependent extinction laws. To avoid the influence of the bias of the extinction coefficients at long wavelengths on the determination of $\Rv$ values, we only used $A_\lambda/\ARP$ with $\lambda < 700 {\rm nm}$.  
The determined total-to-selective extinction ratios are $\Rv=3.40\pm0.07$ and $2.53\pm0.10$ for LMC and SMC, respectively.  
For LMC, our $\Rv$ value is consistent with the average result $\RV=3.41\pm0.06$ from \citet{2003ApJ...594..279G}, which is based on a sample of ten sightlines. 
Different $\RV$ values have also been reported near the 30 Doradus star formation region, such as $2.76\pm0.09$ \citep{2003ApJ...594..279G}, $5.6\pm0.3$ \citep{2014MNRAS.438..513D}, $4.4\pm0.7$ \citep{2014AA...564A..63M}, and $4.5\pm0.2$ \citep{2016MNRAS.455.4373D}.  
For SMC, our $\Rv$ value agrees with the average result of $2.74\pm0.13$ from \citet{2003ApJ...594..279G}, which is for five sightlines around the SMC bar. 
For a small region in the southwest bar of the SMC, a similar value of $R_{\rm 475} = A_{475}/(A_{475}-A_{814}) = 2.65\pm0.11$ was derived by using RC stars \citep{2017ApJ...847..102Y}.	 
Our results are not biased to any specific environment, cover the entire LMC and SMC, and represent the average extinction laws.

\begin{figure}[ht]
\centering
\vspace{0.1in} 
\includegraphics[angle=0,width=3.0in]{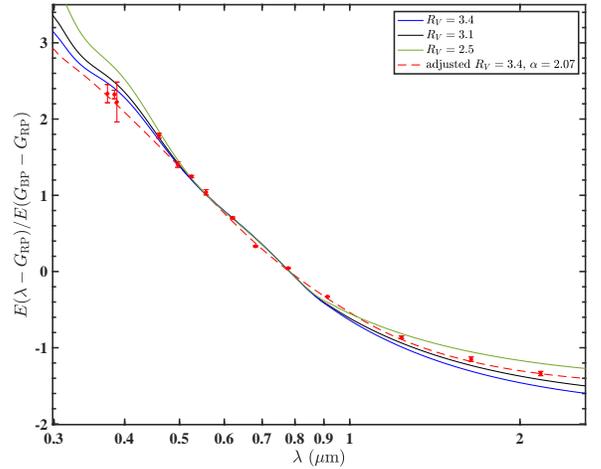}
\vspace{-0.0in}
\caption{\footnotesize
               \label{fig:EElaw_LMC}
The optical--IR reddening curve of the LMC, denoted by color excess ratio $E(\lambda-\GRP)/E(\GBP-\GRP)$ (red filled circles with red error bars).
Red dashed line: our adjusted LMC extinction curve with $\RV = 3.4$.  
The CCM extinction curves with $\Rv$ = 3.4 (blue line), 3.1 (black line), 2.5 (green line) are also shown. 
}
\end{figure}

\subsection{Optical--IR Reddening Curve}

The reddening curve, $\frac{E(\lambda - \GRP)}{E(\GBP - \GRP)}$, is also used to infer the extinction law, especially when the base relative extinction, such as $\frac{\ARP}{E(\GBP - \GRP)}$, $\frac{\ABP}{\ARP}$, is not available. 
In this subsection, we discussed the feasibility of using the reddening curve to infer the extinction law. Figure~\ref{fig:EElaw_LMC} shows the optical--IR reddening curve of the LMC represented by the CER $E(\lambda-\GRP)/E(\GBP-\GRP)$. 
Our adjusted $\RV$ extinction curve and CCM $\Rv$ extinction curves were converted to reddening curves by the equation $E(\lambda-\GRP)/E(\GBP-\GRP)=(A_\lambda/\ARP-1)/(\ABP/\ARP-1)$. The LMC CERs are flatter than the CCM reddening curves but agree well with our adjusted $\RV=3.4$ reddening curve. This also independently proves the reliability of our adjusted extinction law.

Compared to extinction curves (Figure~\ref{fig:ext}), the differences between the different $\RV$ is smaller on the reddening curves (Figure~\ref{fig:EElaw_LMC}).
It can be explained by the fact that the CER is the result of smoothing two relative extinction values. 
In IR bands, the CERs of the black line ($\Rv=3.1$ and $\alpha=1.61$) and the red dashed line ($\Rv=3.4$ and $\alpha=2.07$) are close. It indicates that it is difficult to determine the extinction law, i.e., $\Rv$ or the power-law index $\alpha$, using one or several IR CERs. 
This problem can be reduced by combining optical and IR multi-band CERs. , The $\RV$ can be derived from multi-band CERs if the accuracy of CERs is good enough. Since this approach requires high-precision CERs, the intrinsic color index must be determined from spectroscopic data and even for stars with relatively consistent intrinsic color indices (such as RC stars). Otherwise, small errors in CERs can lead to large uncertainty in the extinction law.

It is also worth emphasizing that we recommend using bands with high photometric accuracy, such as Gaia $\GBP$ and $\GRP$ bands, as the x-axis of the CE--CE diagram to reduce the error of CER caused by the fitting method. More discussions about the effects of CE error on the measurement of the CER was presented in \citet{2019ApJ...877..116W}.

\begin{figure*}[ht]
\centering
\vspace{-0.0in}
\includegraphics[angle=0,width=6.6in]{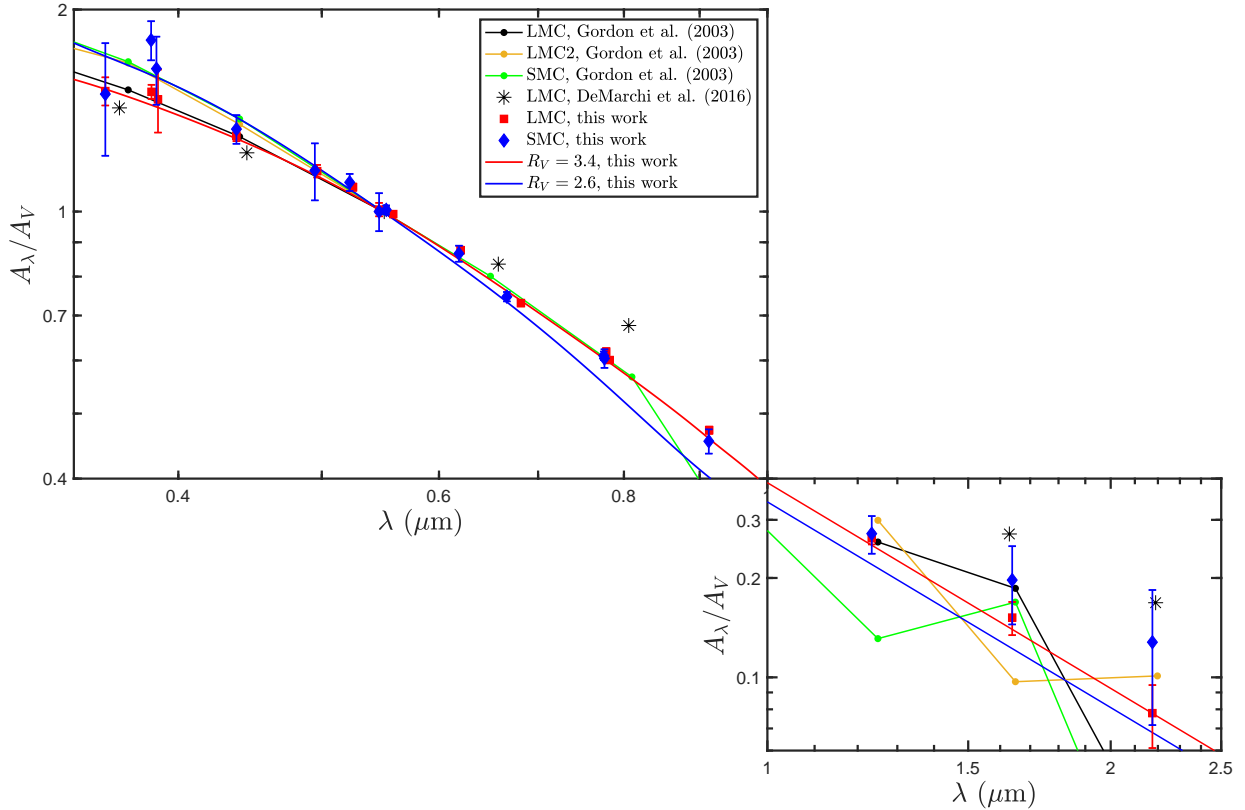}
\vspace{-0.0in}
\caption{\footnotesize
               \label{fig:extlaw}
Comparison of optical to IR multi-band relative extinction $A_{\lambda}/\Av$. 
Our relative extinctions of LMC and SMC are red squares and blue diamonds with error bars, respectively. 
Our adjusted $\RV=3.4$ and $\RV=2.6$ extinction curves are plotted as red and blue lines. 
The extinction curves of \citet{2003ApJ...594..279G} for LMC average, LMC2 supershell, and SMC bar are black, orange, and green lines. 
The LMC 30 Doradus extinction law of \citet{2016MNRAS.455.4373D} are indicated by black asterisks. 
}
\end{figure*}

\section{Discussion}\label{discussion}
\subsection{Comparison with Previous Works}\label{comp}

We compared our determined $\Rv$ values with reported measurements in the literature. 
Table~\ref{tab:Rv} listed some reported $\Rv$ values, the investigated regions, and the adopted tracers. 
Figure~\ref{fig:extlaw} shows the comparison. 
Our LMC extinction are shown as red squares with error bars and can be represented by the $\RV=3.4$ extinction curve (red line). 
Our SMC extinction (blue diamonds with error bars) are significantly steeper than that of LMC and can be represented by the $\RV=2.6$ extinction curve (blue line). 
\citet{2003ApJ...594..279G} measured the LMC (black line) and SMC (green line) average extinction curves.  
Our LMC and SMC extinction curves agree well with the results of \citet{2003ApJ...594..279G} in the optical bands. 
The uncertainties of our relative extinctions are relatively small in most of the bands.
Moreover, our results nicely complement the absence of extinction curves between 0.55 to 1 $\mu$m in \citet{2003ApJ...594..279G}.

In the IR $JH\Ks$ bands, the extinction coefficients in the literature exhibit significant discrepancies, as shown in Figure~\ref{fig:extlaw}. These discrepancies can be explained by the small IR extinction resulting in a large uncertainty in the extinction coefficient.
The extinction in the IR 1.0 to 2.5$\mum$ wavelengths is only 30\% to 8\% of that in the $V$-band. 
For low extinction galaxies like MCs, the accuracy of the average IR extinction law is difficult to improve further. Nevertheless, the IR extinction law is already suitable for most stars in MCs to correct their extinction.

Several studies have investigated the extinction law of the Tarantula Nebula (30 Doradus, a star-forming region) in the LMC. 
For example, \citet{2003ApJ...594..279G} reported $\RV=2.76$ (LMC2, orange line),  \citet{2016MNRAS.455.4373D} derived $\RV= 5.6$ (black asterisks). 
The extinction curve of LMC2 is comparable to that of our SMC. 
In contrast, the extinction curve of \citet{2016MNRAS.455.4373D} is the flattest. 
These works suggest that the extinction law of the LMC may vary with the environment, if the errors in absolute magnitude and intrinsic color are well constrained.

\begin{table*}[ht]                                                                                                    
\begin{center}                                                                                                       
\caption{\label{tab:Rv} Reported $\RV$ Values}                                                                                                             
\begin{tabular}{llll}                                                                                           
\hline \hline
$\RV$ values  &  Region  &  Traces  &  Reference \\
\hline                                                                                                         
${\bf 3.40\pm0.07}$   &   {\bf LMC}   & {\bf 1,073 red supergiant stars} & {\bf This Work} \\
$4.5\pm0.2$   &   LMC 30 Dor & 3,500 red clump stars & \citet{2016MNRAS.455.4373D} \\
$5.6\pm0.3$   &   LMC 30 Dor & 100 red clump stars & \citet{2014MNRAS.438..513D} \\
$4.4\pm0.7$   &   LMC 30 Dor & 83 O- and B-type stars & \citet{2014AA...564A..63M} \\
$3.41\pm0.06$   &   LMC average & 10 O- and B-type stars & \citet{2003ApJ...594..279G} \\   
$2.76\pm0.09$   &   LMC2 supershell (30 Dor) & 9 O- and B-type stars & \citet{2003ApJ...594..279G} \\
$3.16$   &   LMC   &  graphite-silicate model$^*$  & \citet{1992ApJ...395..130P} \\
\hline
${\bf 2.53\pm0.10}$   &   {\bf SMC}   & {\bf 398 red supergiant stars} & {\bf This Work} \\
$2.74\pm0.13$   &   SMC bar & 4 O- and B-type stars & \citet{2003ApJ...594..279G} \\
$2.93$   &   SMC   &  graphite-silicate model$^*$  & \citet{1992ApJ...395..130P}  \\   
\hline
$3.16\pm0.15$   &   Milky Way & 61,111 red clump stars & \citet{2019ApJ...877..116W} \\
$3.32\pm0.18$   &   Milky Way & 37,000 stars & \citet{2016ApJ...821...78S} \\ 
$\sim3.3$ (O-type stars)  and    &   \multirow{2}{*}{Milky Way } & $\sim300$ O-type stars & \multirow{2}{*}{\citet{2016AA...593A.124M}} \\
$\sim3.1$ (classical Cepheids) & & and classical Cepheids & \\
$3.88\pm0.18$ (Westerlund 2) &  open cluster toward & \multirow{2}{*}{ZAMS fitting} & \multirow{2}{*}{\citet{2013AA...555A..50C}} \\
$3.77\pm0.19$ (IC 2581) &  Carina Nebula   &  &  \\
\multirow{2}{*}{2.2 to 4.4}   &   \multirow{2}{*}{within 500 pc of the Sun} & 11,990 O- and B-type stars & \multirow{2}{*}{\citet{2012AstL...38...12G}} \\
  &   & and 30,671 K-type RGB stars &   \\
$3.19\pm0.50$   &   Milky Way & 258 O-type stars & \citet{2003AA...410..905P} \\ 
\hline                                                                                                               
\end{tabular}
\end{center}     
\tablenotetext{}{Note $*$: The extinction results for modeling and determining the $\Rv$ values are summarized in the Table 1 of \citet{1992ApJ...395..130P}  traced by O- and B-type stars.}                                                                                                                                                                                                            
\end{table*}

\begin{figure*}[ht]
\centering
\vspace{-0.0in} 
\includegraphics[angle=0,width=7.2in]{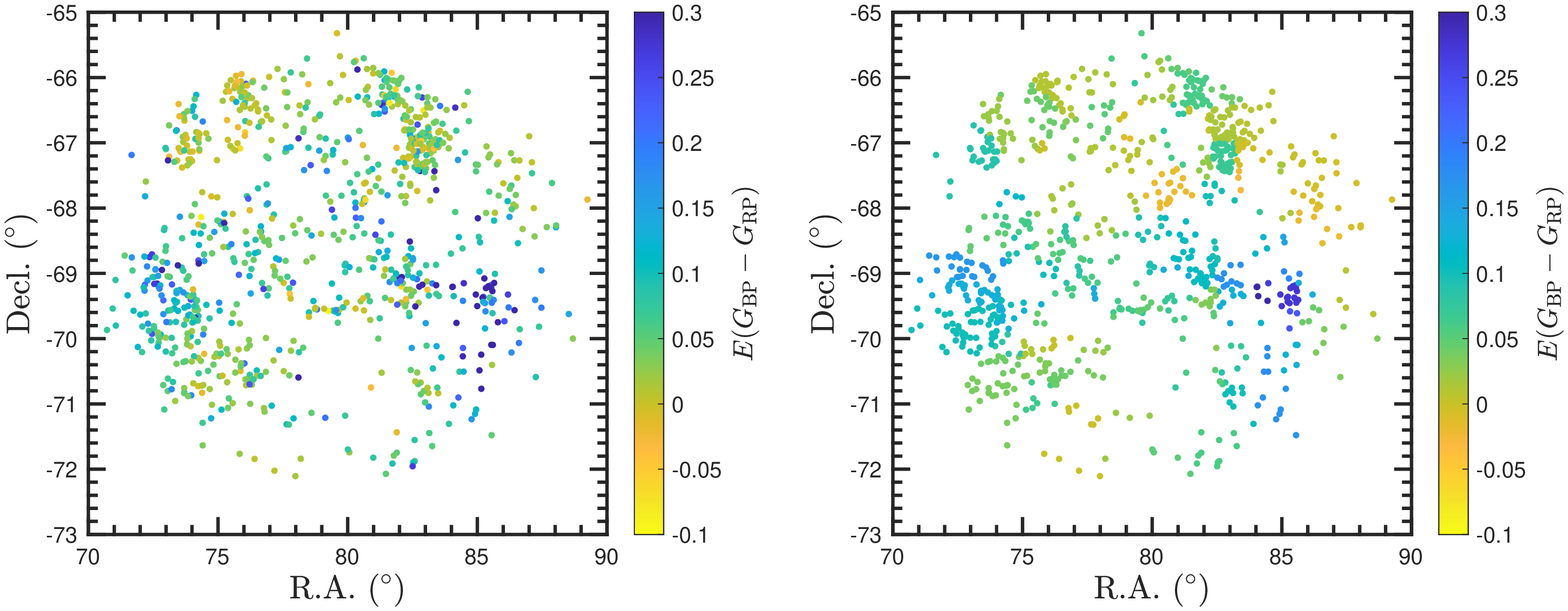}
\vspace{-0.0in}
\caption{\footnotesize
               \label{fig:Emaps}
Comparison of the reddening map based on our LMC RSG stars (left panel) and the Cepheid's reddening map (right panel) from \citet{2019A&A...628A..51J}. The filled circles indicate the position of each RSG star, while colors denote the reddening $E(\GBP-\GRP)$.
}
\end{figure*}

Note that the dispersion of our LMC and SMC $\RV$ values is smaller than that of the Galactic $\Rv$ value from \citet{2019ApJ...877..116W}. 
The error of $\Rv$ mainly comes from $\ARP/E(\GBP-\GRP)$, which determined by the distance and the absolute magnitude. 
\citet{2019ApJ...877..116W} used RC stars with Gaia distance errors less than 10\%, which is larger than the errors of LMC Cepheids' absolute magnitude (0.08 mag).
In addition, compared to the LMC and SMC, the interstellar environment of the Milky Way is more complex, so the dispersion of the Galactic $\RV$ value would be larger.

\subsection{The Effects of Circumstellar Dust}\label{dust}

The mass-loss rate of RSG stars in LMC and SMC has been reported to range from about $10^{-4}$ to $10^{-11}\ {\rm yr}^{-1}$ \citep[e.g.,][]{2005A&A...438..273V, 2012ApJ...753...71R, 2018A&A...609A.114G}. 
The heavy mass loss leads to the formation of the envelope and circumstellar dust. 
We discussed the effects of circumstellar dust on the adoption of RSG stars as interstellar extinction tracers. 
Circumstellar dust absorbs UV/optical photons and re-radiates in the IR.
The dust absorption causes the observed color index of a star to be redder than that without circumstellar dust. 
Although interstellar extinction also leads to a reddening of the observed color index, multi-band interstellar extinction satisfies the extinction law, while circumstellar extinction does not and varies dramatically from star to star.
In addition, the dust emission caused by circumstellar dust leads a steep increase of the IR flux in one to several bands, and the observed IR color indices can deviate significantly from the linear relation between two color indices. 
Therefore, we can check the presence of circumstellar dust in our selected RSG stars based on multi-band color--color diagrams. 
We find that the color--color diagrams of our spectroscopic RSG sample all show a good linearity, indicating that our bands of interest are almost unaffected by circumstellar dust and that the reddening of observed color indices is a result of interstellar extinction.

In addition, IR colors of RSG stars are commonly used to detect circumstellar dust \citep{2005A&A...438..273V, 2010AJ....140..416B, 2011AJ....142..103B, 2018A&A...616A.175Y}. 
Dense circumstellar dust envelopes are more likely to form around RSG stars with large mass-loss rates and low stellar photospheric temperatures \citep{2005A&A...438..273V}.
\citet{2010AJ....140..416B} found that most RSG stars in the SMC have little dust, and only the brightest sources show excess emission could come from circumstellar dust.
This was confirmed by \citet{2011AJ....142..103B}. 
They further found that only the brightest 7\% of LMC RSG stars and 2\% of SMC RSG stars form a significant amount of warm circumstellar dust.
Therefore, after examining the multi-band colors of RSG stars, we excluded RSG stars with circumstellar dust by using $\Ks-W2<0.13$ mag before investigating the interstellar extinction (see Section~\ref{extsample}).

Checking whether what we are studying is interstellar dust can also be confirmed by an external comparison with the extinction map. Considering that RSG stars are young stars, we used the extinction map of Cepheids for comparison. Cepheids are usually considered to have a negligible amount of circumstellar dust. \citet{2019A&A...628A..51J} determined the reddening of 133 segments in the LMC based on $\sim 4500$ Cepheids. The spatial resolution of this reddening map is 1.2 deg$^2$. Based on the positions of our LMC RSG stars, we selected the closest segment to obtain the reddening $E(B-V)$. 
A foreground extinction $E(B-V)=0.05$ mag were subtracted and then $E(B-V)$ was converted into $E(\GBP-\GRP)$ by $E(\GBP-\GRP)/E(B-V)=0.757\pm0.016$ from \citet{2019ApJ...877..116W}. 
Given that we are comparing the internal consistency of the extinction maps, the choice of these parameters will not affect our analysis. 
In Figure~\ref{fig:Emaps}, the $E(\GBP-\GRP)$ of RSG stars (see Section~\ref{CEratio}) is compared to the $E(\GBP-\GRP)$ from Cepheid's reddening map. Overall, the two extinction maps are in excellent agreement, e.g., the maximum extinction occurs near the 30 Doradus region (${\rm R.A.} \sim 85^\circ$ and ${\rm Decl.} \sim -69.5^\circ$). This consistency also suggests that the reddening or extinction we analyzed is dominated by interstellar dust.

\subsection{The Effects of Variability}\label{variable}

The variability of RSG stars is prevalent \citep{1975ApJ...195..137S, 2006MNRAS.372.1721K, 2011ApJ...727...53Y, 2018A&A...616A.175Y, 2019MNRAS.487.4832C, 2020ApJ...898...24R}. 
Compared to AGB stars, RSG stars have smaller amplitudes and reduced variability from the optical to IR bands.
Here, we tried to investigate the impact of variability on the adoption of RSG stars to measure interstellar extinction.

For our LMC RSG sample, we found 360 stars (33\%) with detectable amplitudes out of 1073 RSG stars by analyzing RMSEs estimated by Gaia multi-epoch photometry.
Among them, only 45 stars (4\%) have large amplitude, i.e., $>0.3$ mag in the $G$-band, $>0.2$ mag in the $\GRP$-band, and $>0.1$ mag in the near-IR $\Ks$-band. 
For the $\Ks$-band, the single observation error due to variability is less than 0.03 mag for most RSG stars, which is comparable to the photometric error. 
In optical bands, we used magnitude based on multi-epoch photometry, so the effect of variability on magnitude is limited. 
Moreover, the variation of RSG stars has a much smaller effect on intrinsic colors, since it depends on $\Teff$, which changes very little during light variation. 
For example, the Betelgeuse (a nearby bright RSG star) dims from 0.5 mag to 1.61 mag in the $V$-band, while its $\Teff$ drops only slightly from 3650 K to 3600 K \citep{2020ApJ...891L..37L}. 
This variation in $\Teff$ is comparable to the measurement uncertainty of $\Teff$. 
Our analyses are all based on the intrinsic color of RSG stars, so the effect of variations on the extinction measurements is negligible.

\subsection{Extinction Traced by RGB Stars}\label{RGBext}

RGB stars are often used as extinction tracers in the IR (see Section~\ref{intro}). 
We also tried to use spectroscopic RGB stars to measure the extinction curves of MCs. 
Based on APOGEE parameters, we selected about 3,000 bright LMC RGB stars with $H<13$ mag, $3600 \K \leq \Teff \leq 4200 \K$, and $0\leq \log g \leq 1.2$. 
Then we adopted the same procedure as for RSG stars in Section~\ref{colorexcess} to establish the $\Teff$--$(\lambda-\GRP)_0$ relations and estimate the CERs. 
However, we found that although the number of RGB stars is three times of RSG stars, the RGB stars are almost all located in the low-extinction part of the CE--CE diagram with $E(\GBP-\GRP)<0.2$ mag. 
This leads to significant uncertainty in the CER, i.e., slope, obtained by linearly fitting the CE--CE plot. Currently, the adoption of RGB stars with APOGEE spectra to measure the extinction of MCs is not very suitable.
In this work, we finally chose RSG stars to determine the optical to IR CERs.

\begin{table*}[ht]
\begin{center}
\caption{\label{tab:csst} Predicted LMC and SMC Relative Extinction in CSST Bandpasses}
\vspace{0.1in} 
\hspace{-1.0in}    
\begin{tabular}{lccc}
\hline \hline                                                      
Band ($\lambda$) & $\lambda_{\rm eff}$ ($\mum$)  &   $A_\lambda/A_V$ &   $A_\lambda/A_V$ \\                                                                                              
\hline
& &LMC & SMC  \\                                                                                                     
\hline      
NUV   &  0.2878  &  $1.777$ &  $2.183$ \\
$u$     &  0.3684  &  $1.494$ &  $1.677$ \\
$g$     &  0.4729  &  $1.192$ &  $1.253$ \\   
$r$     &  0.6122  &  $0.863$ &  $0.843$ \\   
$i$     &  0.7579  &  $0.624$ &  $0.575$ \\   
$z$     &  0.8980  &  $0.471$ &  $0.408$ \\  
$y$     &  0.9608  &  $0.418$ &  $0.361$ \\ 
\hline             
\end{tabular}
\end{center}
\end{table*}

\subsection{The Predicted Extinction in CSST Bandpasses} 

Based on the determined extinction law, we predicted the relative extinction in the bandpasses of the China Space Station Telescope (CSST). 
The CSST is a 2-meter aperture survey telescope and a major science project of the China manned space program \citep{Zhan2021CSST}.
The CSST will perform high-resolution, large-area multi-band imaging and slitless spectroscopy surveys. 
It will observe about 17500 deg$^2$ of the sky in the wavelength range of 0.25--1$\mum$, including photometric measurements in seven bands (NUV standing for near-UV, $u$, $g$, $r$, $i$, $z$, $y$) and slitless spectroscopic measurements in three bands (GU, GV, GI). 
The $5 \sigma$ limiting magnitude of the point source can reach 26 (AB mag) or deeper in the $g$ and $r$ bands. 
In the future, CSST data will be suitable for studying the UV-band extinction of the MCs.
Based on our LMC and SMC extinction curves, we determined the relative extinction values $A_\lambda/\AV$ of the CSST bandpasses. 
The extinction coefficients predicted from the effective wavelength of Vega\footnote{The effective wavelengths of the CSST are taken from the SVO Filter Profile Service, http://svo2.cab.inta-csic.es/theory/fps3/.} are listed in Table~\ref{tab:csst}.

\section{Conclusion}\label{conclusion}

We have investigated the optical to IR dust extinction law of LMC and SMC using RSG stars and classical Cepheids as tracers. Cepheids are used to determine the base relative extinction $\ARP/E(\GBP - \GRP)$, while RSG stars are used to determine the multi-band CERs $E(\lambda - \GRP)/E(\GBP - \GRP)$. 
The multi-band photometric data are collected from Gaia, MCPS, SMSS, SMASH, and 2MASS surveys. 
We construct a spectroscopic RSG sample based on APOGEE DR17, with selection criteria including stellar parameters ($\Teff$, $\log g$, [M/H], and [$\alpha$/M]), photometry, and Gaia astrometric data. 
RSG stars are distributed throughout the LMC and SMC, hence, our measurements represent the average extinction laws. 
The main results of this work are as follows.

\begin{enumerate} 
\item 
We present a catalog of spectroscopic RSG stars, including 1073 stars in LMC and 398 stars in SMC. This catalog provides the positions of the RSG stars (R.A., Decl.), the stellar parameters ($\Teff$, $\log g$, and [M/H]), the multi-band photometry with uncertainties (Gaia, MCPS, SMASH, SMSS, and 2MASS), and the derived color excess values $E(\GBP - \GRP)$.  
\item 
We established the $\Teff$--intrinsic color relations and determined the intrinsic colors of RSG stars with $\Teff$ in the range of $3700\K \le \Teff \le 4500\K$ (LMC) and $3900\K \le \Teff \le 4650\K$ (SMC). 
We determined multi-bands CEs and CERs $E(\lambda - \GRP)/E(\GBP - \GRP)$, including one Gaia band ($G$), three Johnson bands ($U, B,V$), two SMASH bands ($u, g$), five SMSS bands ($v, g, r, i, z$), and three 2MASS bands ($J, H, \Ks$).  
\item 
We used classical Cepheids to estimate the base relative extinction $\ARP/E(\GBP-\GRP)$ since the absolute magnitude of Cepheids can be determined more precisely than that of RSG stars. 
The results are $\ARP/E(\GBP-\GRP)=(1.589\pm0.014)$ and $\ARP/E(\GBP-\GRP)=(1.412\pm0.041)$ for LMC and SMC, after a correction for RSG stars.
Combining the $\ARP/E(\GBP-\GRP)$ value with the derived CERs, we obtained the LMC and SMC optical--IR relative extinction $A_\lambda/\ARP$. 
\item
Compared with the CCM model extinction curves, the CCM curves can only well explain the observed extinction in the wavelength range of 300-550 nm. 
In the long-wavelength bands of 550-900 nm, including the Gaia bands, the CCM extinction curve overestimated by up to $10\%$ compared to the observations. 
\item 
To better describe the observed extinction laws of LMC and SMC, we adjusted the $\Rv$-dependent extinction laws. According to our extinction laws, the total-to-selective extinction ratios of LMC and SMC are $\Rv=3.40\pm0.07$ and $\Rv=2.53\pm0.10$. Compared with previous works, our LMC and SMC extinction laws are consistent with \citet{2003ApJ...594..279G} and complement their absence of extinction values between 0.55 to 1 $\mu$m. 
\item 
We showed that the effect of circumstellar dust is small in our RSG sample.
Based on the LMC and SMC extinction laws, we predicted the relative extinctions $A_\lambda/\AV$ and $A_\lambda/\ARP$ for the CSST bandpasses.
\end{enumerate}

\section*{Acknowledgements}
We thank the anonymous referee for very useful comments/suggestions. 
We thank Dr. Ming Yang for very helpful discussions. 
This work is supported by the National Natural Science Foundation of China (NSFC) through the projects 12003046, 12173047, 11903045, 12133002, and 11973001. 
This work is also supported by the National Key Research and Development Program of China, grant 2019YFA0405504 and the science research grants from the China Manned Space Project with No. CMS-CSST-220221-A09. 
S.W. and X.C. acknowledge support from the Youth Innovation Promotion Association of the CAS (grant No. 2022055). 
This work has made use of data from the Gaia, APOGEE, MCPS, SMSS, SMASH, and 2MASS surveys.

This work has made use of data from the European Space Agency (ESA) mission
Gaia (\url{https://www.cosmos.esa.int/gaia}), processed by the Gaia
Data Processing and Analysis Consortium (DPAC,
\url{https://www.cosmos.esa.int/web/gaia/dpac/consortium}). Funding for the DPAC
has been provided by national institutions, in particular the institutions
participating in the Gaia Multilateral Agreement.
APOGEE survey is part of Sloan Digital Sky Survey (SDSS) IV. 
SDSS-IV acknowledges support and resources from the Center for High Performance Computing  at the University of Utah. SDSS-IV is managed by the Astrophysical Research Consortium for the Participating Institutions of the SDSS Collaboration (\url{https://www.sdss.org}). 
This work has made use of data from the SkyMapper Southern Sky survey (SMSS, \url{http://skymapper.anu.edu.au}). 
SkyMapper is owned and operated by The Australian National University's Research School of Astronomy and Astrophysics. 
The survey data were processed and provided by the SkyMapper Team at ANU. 
The Two Micron All Sky Survey is a joint project of the University of Massachusetts and the Infrared Processing and Analysis Center/California Institute of Technology, funded by the NASA and the NSF. 

\bibliography{reference}{}
\bibliographystyle{aasjournal}

\appendix

\setcounter{table}{0}
\renewcommand{\thetable}{A\arabic{table}}
\setcounter{figure}{0}
\renewcommand{\thefigure}{A\arabic{figure}}

\section{Supplementary Figures}

\begin{figure*}[ht]
\centering
\vspace{-0.0in}
\includegraphics[angle=0,width=6.8in]{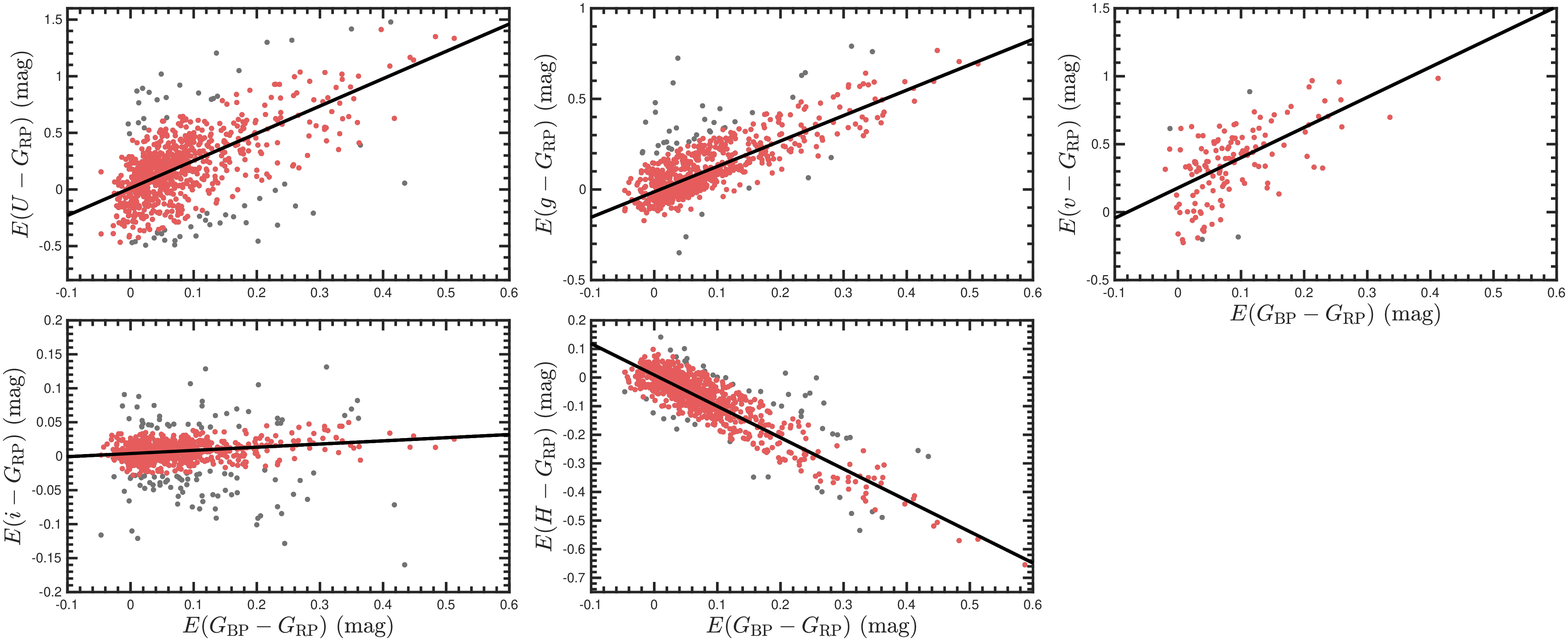}
\caption{\footnotesize
               \label{fig:EE2}
The color excess--color excess diagrams  $E(\GBP - \GRP)$ vs. $E(\lambda - \GRP)$ of RSG stars in the LMC. 
Similar to Figure~\ref{fig:EE1}, but $\lambda$ are $U$ from MCPS, $g$ from SMASH, $v$ and $i$ from SMSS, and $H$ from 2MASS, respectively, from the top left to the bottom right.}
\end{figure*}

\begin{figure*}[ht]
\centering
\vspace{-0.0in}
\includegraphics[angle=0,width=6.8in]{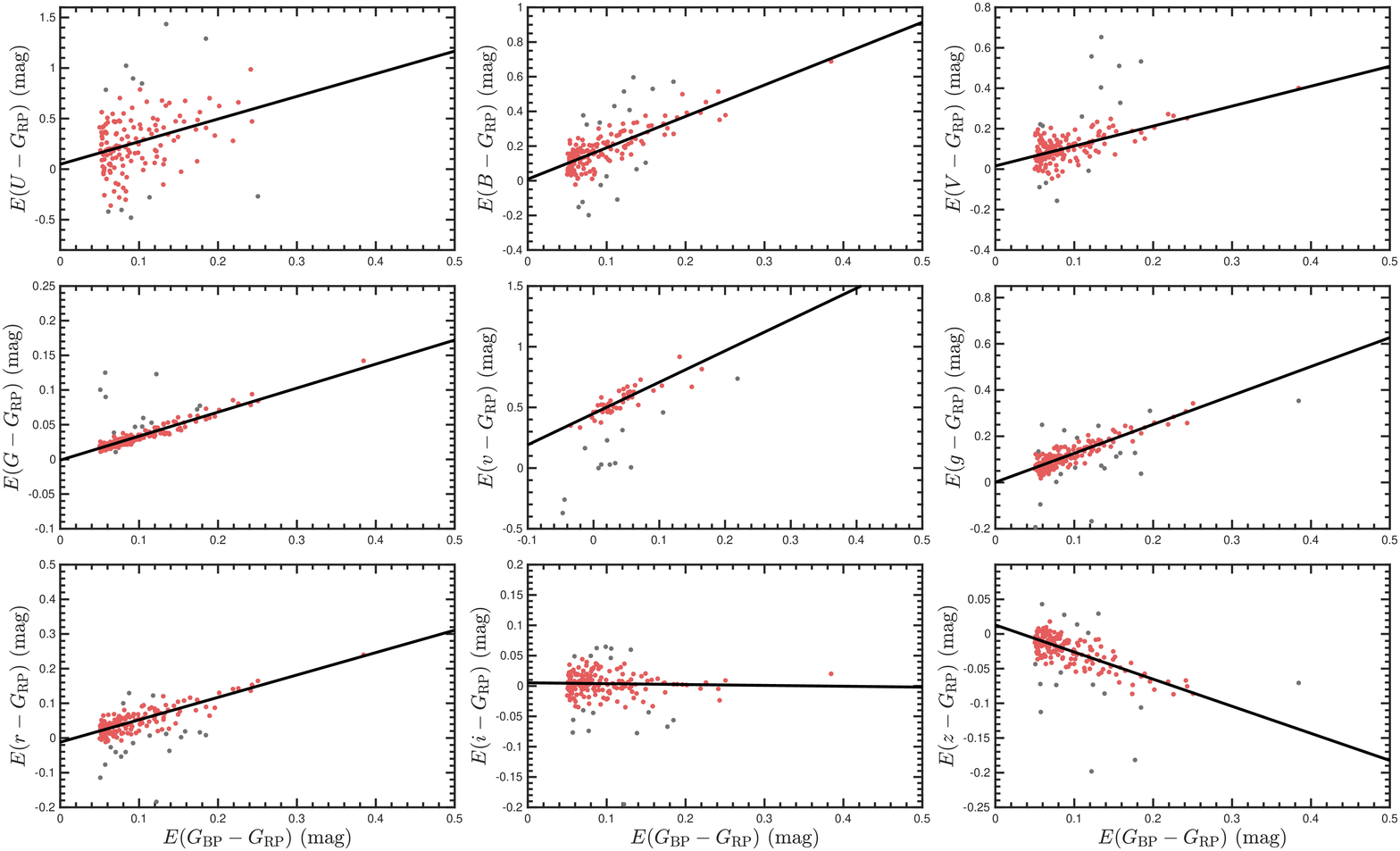}
\vspace{-0.0in}
\caption{\footnotesize
               \label{fig:EE3}
The color excess--color excess diagrams  $E(\GBP - \GRP)$ vs. $E(\lambda - \GRP)$ of RSG stars in the SMC, where  $\lambda$ are  
$U$, $B$, and $V$ from MCPS, $G$ from Gaia
$v$, $g$, $r$, $i$, and $z$ from SMSS, respectively, from the top left to the bottom right.}
\end{figure*}

\begin{figure*}[ht]
\centering
\vspace{-0.0in}
\includegraphics[angle=0,width=6.8in]{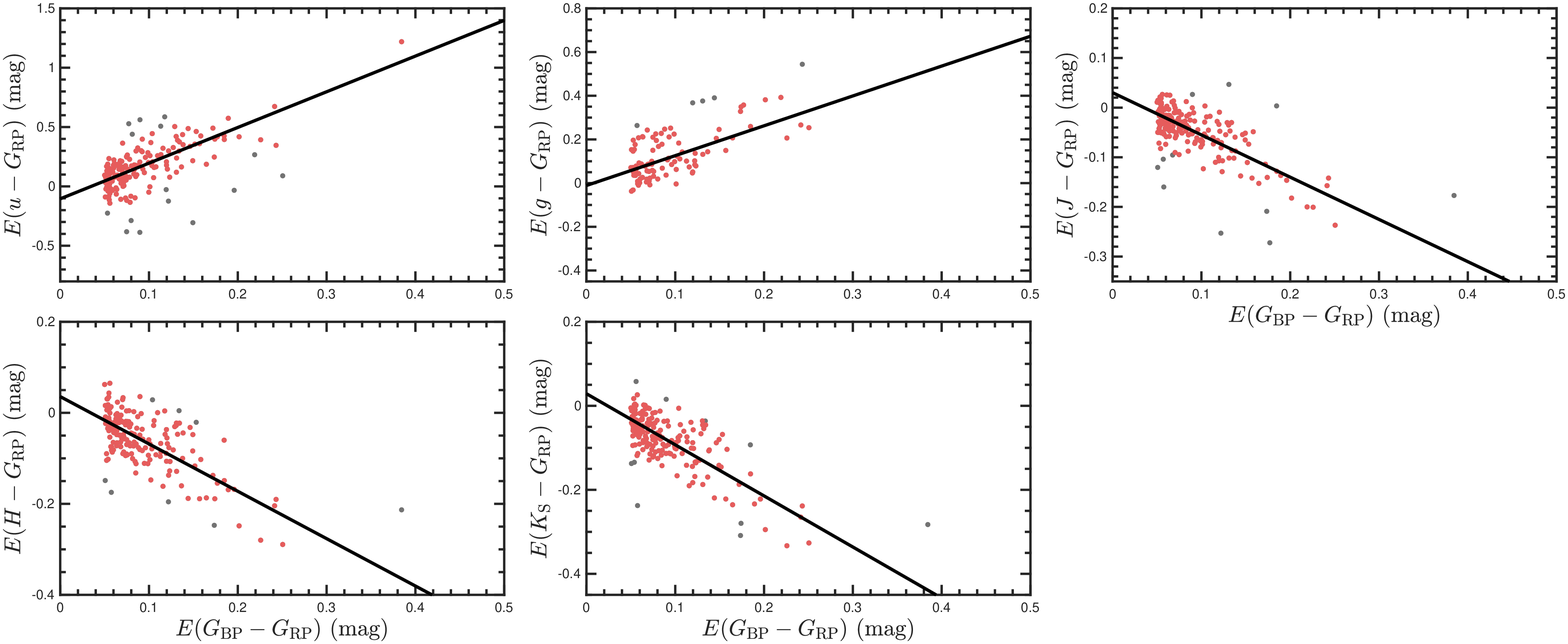}
\caption{\footnotesize
               \label{fig:EE4}
Same color excess--color excess diagrams  $E(\GBP - \GRP)$ vs. $E(\lambda - \GRP)$ as Figure~\ref{fig:EE3}, but $\lambda$ are 
$u$, and $g$ from SMASH, 
$J$, $H$, and $\Ks$ from 2MASS, respectively, from the top left to the bottom right.}
\end{figure*}

\end{document}